\RequirePackage{setspace}
\documentclass[11pt]{article}  
\usepackage{jcappub}

\usepackage{amsfonts}
\usepackage{amssymb}
\usepackage{amsmath}
\usepackage{bm}
\usepackage{leftindex}
\usepackage{graphicx}
\usepackage{rotating}
\usepackage{calligra}
\usepackage{accents}
\usepackage[T1]{fontenc}
\usepackage{scalerel}
\usepackage{nicefrac}

\usepackage[margin=1in]{geometry}
\usepackage{setspace}

\newcommand\omegaLC{{\overset{\circ}{\omega}}{}}
\newcommand\nablaLC{{\overset{\circ}{\nabla}}{}}
\newcommand\DLC{{\overset{\ \circ}{D}}{}}
\newcommand\TLC{{\overset{\ \circ}{T}}{}}
\newcommand\RLC{{\overset{\ \circ}{R}}{}}
\newcommand\QLC{{\overset{\ \circ}{Q}}{}}

\newcommand\scalemath[2]{\scalebox{#1}{\mbox{\ensuremath{\displaystyle #2}}}}
\newcommand\ph{\leftindex_{(p)}{ h}}
\newcommand\pT{\leftindex_{(p)}{ T}}
\newcommand\pbh{\leftindex_{(p)}{ \bar{h}}}

\begin{document}


\title{Teleparallel Geometry with Spherical Symmetry: The diagonal and proper frames.}

\author{R. J.  {van den Hoogen}}
\emailAdd{rvandenh@stfx.ca}
\affiliation{Department of Mathematics and Statistics, St. Francis Xavier University, Antigonish, Nova Scotia, Canada, B2G 2W5}

\author{and H.  {Forance}}
\emailAdd{x2021cij@stfx.ca}

\date{\today}


\abstract{
We present the proper co-frame and its corresponding (diagonal) co-frame/spin connection pair for spherically symmetric geometries which can be used as an initial ansatz in any theory of teleparallel gravity.  The Lorentz transformation facilitating the move from one co-frame to the other is also presented in factored form. The factored form also illustrates the nature of the two degrees of freedom found in the spin connection. The choice of coordinates in restricting the number of arbitrary functions is also presented. Beginning with a thorough presentation of teleparallel gravity using the metric affine gauge theory approach and concentrating on $f(T)$ teleparallel gravity, we express the field equations in the diagonal co-frame. We argue that the choice of diagonal co-frame may be more advantageous over the proper co-frame choice.  Finally, assuming one additional symmetry, we restrict ourselves to the Kantowski-Sachs teleparallel geometries, and determine some solutions.
}

\maketitle


\section{Introduction}

\subsection{Preface}
Using geometrical symmetries to construct gravitational and cosmological models in Einstein's theory of General Relativity (GR) is well developed and has a long history\cite{Stephani:2003tm,hall2004symmetries}. In the early history of GR, a solution describing the geometry of a static spherically symmetric vacuum space-time was developed by Schwarschild. This Schwarschild solution to the Einstein field equations (FEs) remains among one of the most important and widely used solutions to date because of its success in explaining various deviations from Newtonian dynamics observed in the solar system to a tremendous degree of accuracy. Further, spherically symmetric solutions of GR containing a perfect fluid matter source are also important in constructing and understanding relativistic effects in some simple astrophysical and stellar models \cite{Stephani:2004ud,Groen:2007zz,Padmanabhan:2010zzb}. Additionally, spherically symmetric GR models are employed in cosmological applications. For example, the LeMaitre-Tolman-Bondi, Kantowski-Sachs, and Friedmann-Robertson-Walker models provide useful toy models to test different cosmological ideas and observations \cite{Stephani:2003tm,Krasinski:1997yxj}.

It is therefore important that any alternative theory of gravity have comparable analogues to these spherically symmetric GR solutions. GR is a theory of gravity that uses a pseudo-Riemannian geometrical framework in which the connection is metric compatible, has zero torsion and non-zero curvature.  If one moves away from pseudo-Riemannian geometries to teleparallel geometries which have a metric-compatible connection with a non-trivial torsion and a trivial curvature, one can construct alternative theories of gravity, collectively called Teleparallel Gravity.  Interestingly, an analogue to GR known as the Teleparallel Equivalent to GR (TEGR) can also be constructed \cite{Aldrovandi_Pereira2013,Maluf2013,Cai_2015,Bahamonde:2021gfp,Krssak:2018ywd}. TEGR is dynamically equivalent to GR yet the FEs in this theory are of second order, avoiding the issues that come with theories having FEs of higher order. As in GR, the Lagrangian for TEGR depends linearly on a scalar $T$ produced from contractions of the torsion tensor.  A popular generalization of TEGR are the $f(T)$ class of teleparallel theories of gravity \cite{Ferraro:2006jd,Krssak_Saridakis2015,Bahamonde:2021gfp,Krssak:2018ywd}, New General Relativity \cite{Hayashi:1979qx,Bahamonde:2021gfp} and generalizations thereof \cite{Bahamonde:2017wwk,Bahamonde:2021gfp}.

Previous to the popularization of the fully covariant approach to $f(T)$ teleparallel gravity \cite{Krssak:2018ywd,Krssak_Saridakis2015,Krssak_Pereira2015} the literature was plagued with confusion about whether a frame in which inertial effects are absent can be found without violating Lorentz invariance. These so called \emph{proper frames} mathematically correspond to those frames for which the associated spin connection is trivial. These proper frames are devoid of inertial effects, allowing for greater focus on the effects of gravity. This way of separating inertial and gravitational effects is one of the main theoretical advantages of teleparallel gravity \cite{Aldrovandi_Pereira2013}. At the introduction of $f(T)$ teleparallel gravity (and other classes of teleparallel gravity), a popular approach was to simply assume a trivial spin connection, leading to situations where two frames related by a local Lorentz transformation may not both solve the FEs thus violating Lorentz invariance \cite{Li_Sotiriou_Barrow2010}.  Some authors in the past have employed a constructive approach of finding a frame for which the corresponding trivial spin connection yields field equations in $f(T)$ teleparallel gravity which do not limit the function $f(T)$ to be linear in $T$, labelling those frames which work as `good tetrads' and those that do not, `bad tetrads' \cite{Tamanini_Boehmer2012}. However the idea of good and bad tetrads in teleparallel gravity is not based on symmetry assumptions, but on solutions to the antisymmetric part of the field equations.

The metric affine gauge (MAG) approach resolves some of these issues at the cost of relaxing the condition of a trivial spin connection \cite{Hehl_McCrea_Mielke_Neeman1995}. No assumptions about the geometry are made apriori; instead Lagrange multipliers are used to constrain the non-metricity and curvature to be zero \cite{Obukhov_Pereira2003}. This construction, summarized in Sections \ref{MAG_approach}  and  \ref{tele-mag},  yields a fully covariant framework for teleparallel gravity, including $f(T)$ teleparallel gravity.  In addition to providing a self-contained exposition of teleparallel gravity in the MAG framework, we carefully illustrate how the translational coupling prescription \cite{Aldrovandi_Pereira2013} yields a matter source that is minimally coupled to the geometry thereby respecting the equivalence principle.  However, even within the MAG approach to teleparallel gravity, given a set of symmetries, for example spherical symmetry, the issue remains of first determining either the proper frame, or a frame/spin-connection pair, which satisfy the symmetry requirements.

In Riemannian and pseudo-Riemannian geometries, the geometrical setting for gravitational theories such as GR, symmetries are defined as isometries of the metric \cite{Stephani:2003tm}. An isometry being a differentiable mapping of a manifold into itself preserving the notion of lengths and angles as determined by the metric. The corresponding vector fields that generate these symmetries are known as Killing vectors and together with the metric satisfy Killing's equation.  Since the Levi-Civita connection is computed directly from the metric, it too is invariant under the same symmetry requirements.  That is, the isometry of the metric imposes symmetry conditions on the connection.

However, in Riemann-Cartan geometries in which there exists an independent connection, one necessarily requires an additional assumption about how any symmetry assumption limits this independent connection. That is, we must define an affine symmetry. In teleparallel geometries, (a subset of Riemann-Cartan geometries) where the basis frame (or basis co-frame) can be considered the primary object of interest alongside the connection, further analysis is required to understand how affine frame symmetries can be imposed on these teleparallel geometries \cite{FJChinea_1988,Fonseca-Neto:1992xln,fon2002,McNutt:2023nxm}. When approaching this problem one must consider the possibility of non-trivial isotropy in the desired geometry to ensure the resulting frame-spin connection pair obeys such conditions\cite{McNutt:2023nxm}. A more concrete difference perhaps, the group of affine frame symmetries may be smaller than that of the group of isometries. For example, it has been shown that the only teleparallel geometry in four dimensions that admits a ten dimensional group of affine frame symmetries, is the trivial Minkowski geometry\cite{Coley:2019zld} where the torsion is identically zero. Contrasting this observation with Riemannian geometries in four dimensions where DeSitter, Minkowski and anti-DeSitter geometries all admit ten dimensional groups of isometries.

In \cite{McNutt:2023nxm} an algorithm to determine the most general Riemann-Cartan geometry which admits a given symmetry group was presented through the use of invariantly defined frames resulting from an application of the Cartan-Karlhede algorithm \cite{Fonseca-Neto:1992xln,Coley:2019zld}.  In this way, the number of arbitrary functions defining the resultant geometry described by the frame and connection is minimized.  In particular, because of its importance in astrophysical and cosmological applications, using the procedure in \cite{McNutt:2023nxm}, the most general spherically symmetric (non-static) frame/spin-connection pair have been determined for teleparallel geometries. The resultant frame is diagonal and the corresponding spin connection is non-trivial.  This spherically symmetric (diagonal) frame/spin-connection pair has been used to construct various spherically symmetric models in $f(T)$ teleparallel gravity \cite{Coley:2022qug,Coley:2023kyg,Coley:2023dbg,Coley:2024tqe,Landry:2024dzq,Landry:2024pxm}. In Section \ref{section-diagonal-frame} we construct the corresponding proper frame associated with the (diagonal) frame/spin-connection pair presented in \cite{McNutt:2023nxm}.  We also provide a decomposition of the corresponding Lorentz transformation in terms of simple rotations and boosts.

An alternative and complementary approach to determine the spherically symmetric frame in metric-affine geometries including teleparallel geometries has been developed by Hohmann and collaborators \cite{Hohmann:2019nat,Hohmann:2019fvf,Bahamonde:2021gfp}. One of the differences between the end result of the two different approaches is that the approach of \cite{McNutt:2023nxm} yields a simple diagonal frame and its corresponding non-trivial spin connection, whereas the approach of \cite{Hohmann:2019nat,Hohmann:2019fvf,Bahamonde:2021gfp} yields a non-diagonal frame with a trivial connection, i.e., the proper frame. In Section \ref{Compare-Hohmann} it is shown explicitly that the two results are equivalent and the transformation that relates them is given.

In Section \ref{section5} using the MAG approach illustrated earlier, we present the covariant $f(T)$ teleparallel gravity field equations for a general spherically symmetric geometry in the diagonal frame. We also illustrate, independent of the field equations, with a simple example that using the proper frame may lead to some computational difficulties. This provides some evidence that working in the diagonal frame and its corresponding spin connection might be computationally more efficient when investigating teleparallel geometries.

Unfortunately, there are some errors in the current teleparallel gravity literature \cite{Rodrigues:2012qua,Rodrigues:2014xam,Cai_2015} as it pertains to the spherically symmetric Kantowski-Sachs geometries in teleparallel gravity. Since the Kantowski-Sachs model necessarily has a 3 dimensional spatial hyper-surface with positive spatial curvature, any line element of the form
$$ds^2 = -dt^2+a^2(t)\,dx^2+e^{-2\alpha x}b^2(t)\,dy^2+c^2(t)\, dz^2 $$
where $\alpha=0$ and $b(t)=c(t)$ is clearly Bianchi I which has flat spatial hyper-surfaces and is in no way Kantowski-Sachs. It can only be the asymptotic zero spatial curvature limit of the Kantowski-Sachs model. In Section \ref{Kantowski-Sachs} we employ the correct geometrical ansatz for the Kantowski-Sachs geometries in teleparallel gravity and present the covariant $f(T)$ teleparallel gravity field equations. We also provide some simple solutions to the field equations, complementing the work of Landry \cite{Landry:2024pxm}.

\subsection{Notation}

The notation employed uses Greek indices $(\mu,\nu,\dots)$ to represent space-time coordinate indices, while Latin indices $(a,b,\dots)$, represent frame or tangent-space indices.  Upper case Latin indices $\{I,J,\dots\}$ represent the restricted set of tangent space indices $\{1,2,3\}$. Particular values of tangent or cotangent space indices are indicated with a hat. The wedge $\wedge$ denotes the exterior product, while the symbol $\rfloor$ denotes the interior product of a vector and a $p$-form.  Round brackets surrounding indices represents symmetrization, while square brackets represents anti-symmetrization. Any underlined index is not included in the symmetrization.
The metric signature is assumed to be $(-,+,+,+)$.
The frame basis is denoted as $e_a$ with the corresponding co-frame basis $h^a$ where $e_a\rfloor h^b = \delta^a_b$.
A proper co-frame basis is denoted as $\ph^a$.  The exterior covariant derivative, denoted by $D$, of a matrix-valued p-form $U^a_{~b}$ with respect to the general linear spin connection $\omega^a_{~b}$ is
\begin{equation}
DU^a_{~b}:=dU^a_{~b}+\omega^a_{~c}\wedge U^c_{~b}-\omega^c_{~b}\wedge U^a_{~c}
\end{equation}
where $d$ indicates the exterior derivative.
Quantities having a circle above the symbol are calculated using the Levi-Civita spin connection, denoted as $\omegaLC^a_{\phantom{a}b}$ such that
\begin{equation}
\DLC U^a_{~b}:=dU^a_{~b}+\omegaLC^a_{~c}\wedge U^c_{~b}-\omegaLC^c_{~b}\wedge U^a_{~c}.
\end{equation}
Further, an inertial holonomic co-frame and quantities derived from it are indicated using a tilde, for example, the co-frame $\tilde{h}^a$ has a torsion $\tilde{T}^a$ which is identically zero.
We define the volume four form $\eta={}^*(1)$ as the dual to the unit zero-form and through continued inner products with frame $e_a$ we are able to define $\eta_a=e_a\rfloor\eta = {}^*h_a$, $\eta_{ab}=e_a\rfloor e_b\rfloor \eta = {}^*(h_a\wedge h_b)$ and $\eta_{abc}=e_a\rfloor e_b\rfloor e_c\rfloor \eta = {}^*(h_a\wedge h_b \wedge h_c)$ where the one form $h_a=g_{ab}h^b$.


\section{Metric Affine Gauge Theory Approach} \label{MAG_approach}

The Metric Affine Gauge (MAG) theory framework has been presented in an excellent exposition by Hehl et al. \cite{Hehl_McCrea_Mielke_Neeman1995}.  Below, we summarize the pertinent points in a self-consistent manner applicable to teleparallel gravity and the assumed sign convention so as to make this paper self-contained.  The reader is encouraged to consult \cite{Hehl_McCrea_Mielke_Neeman1995} for complete details.

\subsection{Co-frame, Metric, and Connection}

Consider a $4$-dimensional differentiable manifold $M$ with coordinates $\{x^\mu\}$. Since $M$ is differentiable, at every point, there exists four linearly independent vectors $e_a$ tangent to the manifold at that point. Further, these four vector fields also define the dual or co-frame field $h^a$ uniquely through the relation $e_b\rfloor h^a = \delta^a_b$. Lengths and angles are measured via the additional assumption of a symmetric metric field $\mathbf{g}$ on the manifold $M$, having components $g_{ab}=\mathbf{g}(e_a,e_b)$. Further, through the assumption of a linear affine connection $\bm{\omega}$, represented as a matrix-valued one-form $\omega^{a}_{\phantom{a}b}$ or spin connection, one is able to define a covariant differentiation process.  The geometrical quantities $\{g_{ab}, h^a,\omega^{a}_{\phantom{a}b}\}$ are independent in general,  but with additional geometrical or physical assumptions, interesting inter-relationships between $\{g_{ab}, h^a, \omega^{a}_{\phantom{a}b}\}$ appear.

For example, it is widely accepted that the laws of gravitational physics be invariant under arbitrary changes of coordinates ({\em General Covariance}) and be invariant under arbitrary changes in the frame of reference ({\em Principle of Relativity}) \cite{Ortin:2004ms,Hehl_McCrea_Mielke_Neeman1995}.  These two conditions imply that the field equations derived from the Lagrangian describing some gravitational theory transform covariantly under general coordinate transformations and transform covariantly under local linear transformations of the frame in four dimensions, i.e., $GL(4,\mathbb{R})$ gauge transformations of the co-frame, metric, and affine connection fields.

\subsection{Field Strengths and Bianchi Identities}

The field strengths for the metric $g_{ab}$, co-frame $h^a$, and spin connection $\omega^{a}_{\phantom{a}b}$ are respectively \cite{Hehl_McCrea_Mielke_Neeman1995},
\begin{subequations}
\begin{align}
\text{Non-metricity one-form:}\qquad  & Q_{ab}:=-Dg_{ab}=-dg_{ab}+2\omega^c_{\phantom{a}(a}g^{\vphantom{a}}_{b)c},\label{nonmetricity}\\
\text{Torsion two-form:}\qquad  & T^a:=Dh^a=dh^a+\omega^a_{\phantom{a}b}\wedge h^b,\label{torsion} \\
\text{Curvature two-form:}\qquad  & R^a_{\phantom{a}b}:=d\omega^a_{\phantom{a}b}+\omega^a_{\phantom{a}c}\wedge\omega^c_{\phantom{a}b}.\label{curvature}
\end{align}
\end{subequations}
The associated Bianchi identities are
\begin{subequations}
\begin{align}
DQ_{ab}             &= 2R_{(ab)},\label{Bianchi1}\\
DT^a                &= R^{a}_{\phantom{a}c} \wedge h^c, \label{Bianchi2}\\
DR^a_{\phantom{a}b} &= 0\label{Bianchi3},
\end{align}
\end{subequations}
which yield additional differential constraints. We note here that in teleparallel theories of gravity in which the curvature and non-metricity are zero, the Bianchi identities are identically satisfied, which is not the case in say  Riemannian theories of gravity like GR in which \eqref{Bianchi2} imposes a non-trivial constraint on the Riemannian curvature.

\subsection{Anholonimity, Levi-Civita Connection and the Contorsion}

The spin-connection one-form $\omega^a_{~b}$ can be decomposed into Riemannian and non-Riemannian parts using equations \eqref{nonmetricity}, \eqref{torsion} and the anholonimity two-form $C^a$,
\begin{equation}
C^a:=dh^a.
\end{equation}
Using the metric to move both indices down, $\omega_{ab}:=g_{ac}\omega^c_{~b}$ it can be shown that\footnote{Due to the convention of placing indices in the definition of the covariant derivative, there are a number of sign differences between what  is presented here and what is in \cite{Hehl_McCrea_Mielke_Neeman1995}.}
\begin{equation}\label{conto_LC}
\omega_{ab} = \omegaLC_{ab} + K_{ab} +\frac{1}{2}Q_{ab}-(e_{[a}\rfloor Q_{b]c})\,h^c,
\end{equation}
where
\begin{equation}\label{conn_LC}
\omegaLC_{ab} := \frac{1}{2}dg_{ab}-(e_{[a}\rfloor dg_{b]c})\,h^c-e_{[a}\rfloor C_{b]} +\frac{1}{2}(e_a\rfloor e_b\rfloor C_c)\,h^c,
\end{equation}
is the Riemannian part of the spin connection, which we indicate with a circle above the symbol. Since it corresponds to the Levi-Civita connection of the metric, we also call $\omegaLC^a_{~b}$ the Levi-Civita spin connection one-form.  The contorsion one-form $K^a_{~b}$ where $K_{ab}=-K_{ba}$ is defined implicitly from the torsion two-form via
\begin{equation}\label{def_conto}
T^a:=K^a_{~b}\wedge h^b,
\end{equation}
and since $K_{ab}$ is antisymmetric in its indices, it can be inverted,
\begin{equation}
K_{ab}:=e_{[a}\rfloor T_{b]} -\frac{1}{2}(e_a\rfloor e_b\rfloor T_c)\,h^c,
\end{equation}
yielding an explicit expression for the contorsion.

%
%
%

\subsection{The Matter and Gauge Field Lagrangians}

The matter fields can be described using scalar, tensor or spinor-valued forms of any rank. Here, it is sufficient for our purposes to consider a scalar field $\Phi$ and a vector valued $1$-form field $\Psi^a$. The $4$-form matter field Lagrangian, $L_{Matt}$, is a function of the field variables $g_{ab}$, $h^a$, $\omega^a_{\phantom{a}b}$, $\Phi$, $\Psi^a$ and their corresponding derivatives $dg_{ab}$, $dh^a$, $d\omega^a_{\phantom{a}b}$, $d\Phi$, $d\Psi^a$. However, the {\em Principle of Relativity}, implies that the spin connection, $\omega^a_{\phantom{a}b}$ can only enter the matter field Lagrangian via covariant derivatives of the field variables,  hence
\begin{equation}
L_{Matt}:=L_{Matt}(g_{ab},Q_{ab},h^a,T^a,R^a_{\phantom{a}b},\Phi,d\Phi,\Psi^a,D\Psi^a).\label{Matter_Lagrangian}
\end{equation}
where $D$ represents the covariant exterior derivative. We shall argue later that \eqref{Matter_Lagrangian} in its current form assumes a non-minimal coupling between the matter fields and the geometry. A minimal coupling prescription between the matter and the geometry for teleparallel gravity is presented in Section \ref{minimal-coupling}.

Similarly, with the assumption of the {\em Principle of Relativity}, the $4$-form gauge field Lagrangian, $V_{gauge}$, can be only a function of the field variables $g_{ab},h^a,\omega^a_{\phantom{a}b}$ and their corresponding covariant derivatives, hence
\begin{equation}
V_{gauge}:=V_{gauge}(g_{ab},Q_{ab},h^a,T^a,R^a_{\phantom{a}b}).
\end{equation}

The definitions of the metrical energy momentum, canonical energy momentum, and the hyper-momentum matter currents \cite{Hehl_McCrea_Mielke_Neeman1995} are respectively
\begin{align}
\sigma^{ab} &:= 2\frac{\delta L_{Matt}}{\delta g_{ab}}=2\frac{\partial L_{Matt}}{\partial g_{ab}}+2D\left(\frac{\partial L_{Matt}}{\partial Q_{ab}}\right), \\
\Sigma_a &:= \frac{\delta L_{Matt}}{\delta h^a}=\frac{\partial L_{Matt}}{\partial h^a}+D\left(\frac{\partial L_{Matt}}{\partial T^a}\right), \\
\Delta^a_{\phantom{a}b}&:= \frac{\delta L_{Matt}}{\delta \omega^b_{~a}}=
2g_{bc}\frac{\partial L_{Matt}}{\partial Q_{ac}}+h^a\wedge \frac{\partial L_{Matt}}{\partial T^b}
+D\left(\frac{\partial L_{Matt}}{\partial R^b_{~a}}\right) +\Psi^a \wedge \frac{\partial L_{Matt}}{\partial(D\Psi^b)}.\label{hyper-momentum}
\end{align}

The gauge field momenta \cite{Hehl_McCrea_Mielke_Neeman1995} are
\begin{align}
M^{ab} &:= 2\frac{\partial V_{gauge}}{\partial dg_{ab}}
         =-2\frac{\partial V_{gauge}}{\partial Q_{ab}},\label{gauge_FM1}\\
H_a &:=-\frac{\partial V_{gauge}}{\partial dh^a}
      =-\frac{\partial V_{gauge}}{\partial T^a},\label{gauge_FM2}\\
H^{a}_{\phantom{a}b}&:=-\frac{\partial V_{gauge}}{\partial (d\omega^b_{\phantom{a}a})}
      =-\frac{\partial V_{gauge}}{\partial R^b_{\phantom{a}a}}.\label{gauge_FM3}
\end{align}
The associated metrical energy momentum, canonical energy momentum and the hyper-momentum  of the gauge fields \cite{Hehl_McCrea_Mielke_Neeman1995} are respectively
\begin{align}
m^{ab} &:= 2\frac{\partial V_{gauge}}{\partial g_{ab}}, \label{gauge_EM1}\\
E_a&:=\frac{\partial V_{gauge}}{\partial h^a},\label{gauge_EM2}\\
E^a_{\phantom{a}b}&:=\frac{\partial V_{gauge}}{\partial \omega^b_{\phantom{a}a}}.\label{gauge_EM3}
\end{align}


\subsection{Noether Identities}
If a Lagrangian is invariant under a set of continuous symmetry transformations then Noether's theorem concludes that there exists a set of differential relations between the different variational derivatives.
Since we have assumed {\em General Covariance} and the {\em Principle of Relativity}, we will obtain two differential relations.  For the gauge fields, the Noether identities associated with {\em General Covariance} and the {\em Principle of Relativity} yield the following two expressions respectively \cite{Hehl_McCrea_Mielke_Neeman1995,Hecht:1992xn}
\begin{align}
D\frac{\delta V_{gauge}}{\delta h^a} &= (e_a\rfloor T^b) \wedge \frac{\delta V_{gauge}}{\delta h^b} + (e_a\rfloor R^b_{~c}) \wedge\frac{\delta V_{gauge}}{\delta \omega^b_{~c}} -(e_a\rfloor Q_{bc}) \frac{\delta V_{gauge}}{\delta g_{bc}},
\\
D\frac{\delta V_{gauge}}{\delta \omega^b_{~a}} &= 2g_{bc}\frac{\delta V_{gauge}}{\delta g_{ac}} - h^a\wedge \frac{\delta V_{gauge}}{\delta h^b},\label{noether-g2}
\end{align}
and an expression for the canonical energy momentum for the gauge fields
\begin{align}
\frac{\partial V_{gauge}}{\partial h^a} & = e_a\rfloor V_{gauge} - (e_a\rfloor T^b) \wedge \frac{\partial V_{gauge}}{\partial T^b} - (e_a\rfloor R^b_{~c}) \wedge\frac{\partial V_{gauge}}{\partial R^b_{~c}} -(e_a\rfloor Q_{bc}) \frac{\partial V_{gauge}}{\partial Q_{bc}}.
\end{align}

Similarly, assuming the field equation for the matter fields are satisfied (i.e., on shell),
\begin{equation}
\frac{\delta L_{Matt}}{\delta \Phi}=0 \text{\qquad and \qquad}\frac{\delta L_{Matt}}{\delta \Psi^a}=0,
\end{equation}
the Noether identities for the matter currents associated with {\em General Covariance} and the {\em Principle of Relativity}, yield the following expressions respectively
\begin{align}
D\Sigma_a &= (e_a\rfloor T^b)\wedge \Sigma_b + (e_a\rfloor R^c_{~b})\wedge \Delta^b_{~c} -\frac{1}{2} (e_a\rfloor Q_{bc})\sigma^{bc},\label{noether-m1}\\
D\Delta^a_{~b} &=  g_{bc}\sigma^{ac}-h^a\wedge\Sigma_b,\label{noether-m2}
\end{align}
and an expression for the canonical energy momentum for the matter fields
\begin{align}
\Sigma_a & = e_a\rfloor L_{Matt} - (e_a\rfloor T^b) \wedge \frac{\partial L_{Matt}}{\partial T^b} - (e_a\rfloor R^b_{~c}) \wedge\frac{\partial L_{Matt}}{\partial R^b_{~c}} -(e_a\rfloor Q_{bc}) \frac{\partial L_{Matt}}{\partial Q_{bc}}\nonumber\\
&\quad +D\frac{\partial L_{Matt}}{\partial T^a} -(e_a \rfloor D\Phi)  \frac{\partial L_{Matt}}{\partial D\Phi}
-(e_a \rfloor D\Psi^a) \wedge \frac{\partial L_{Matt}}{\partial D\Psi^a}-(e_a\rfloor \Psi^a)\wedge \frac{\partial L_{Matt}}{\partial \Psi^a}.
\end{align}

\section{Teleparallel Gravity using the MAG Framework} \label{tele-mag}

\subsection{The Action for Teleparallel Gravity}

In any theory of teleparallel gravity, the non-metricity and curvature are identically zero.  Therefore, the matter field and gauge field Lagrangians for teleparallel gravity are of the form
\begin{eqnarray}
L_{Matt}&=&L_{Matt}(g_{ab},h^a,T^a,\Phi,d\Phi,\Psi^a,D\Psi^a), \label{restrict_L}\\
V_{Tele}&=&V_{gauge}(g_{ab},h^a,T^a) +\frac{1}{2}Q_{ab}\wedge\mu^{ab} +R^{a}_{\phantom{a}{b}}\wedge \nu_a^{\phantom{a}{b}}, \label{restrict_V}
\end{eqnarray}
where we have introduced the Lagrange multiplier $\mu^{ab}$ as a symmetric $3$ form and the Lagrange multiplier $\nu_a^{\phantom{a}b}$ as a $2$ form.  The nature of the matter Lagrangian density will be discussed in more detail in Section \ref{source coupling} where we provide further details on how the gravitational field is coupled to the matter fields.

The field equations for teleparallel gravity can be obtained by varying the constrained action
\begin{equation}
S=\int\left(L_{Matt}+V_{Tele}\right),\label{action2}
\end{equation}
with respect to $\mu^{ab},\nu_a^{\phantom{a}b},g_{ab},h^a,\omega^a_{\phantom{a}b},\Phi$, and $\Psi^a$ \cite{Hehl_McCrea_Mielke_Neeman1995,Obukhov_Pereira2003}.


\subsection{Field Equations}

Variation of the action \eqref{action2} with respect to the $\mu^{ab},\nu_a^{\phantom{a}b},g_{ab},h^a,\omega^a_{\phantom{a}b},\Phi$, and $\Psi^a $ yields
\begin{eqnarray}
Q_{ab}=0, &&\qquad\mbox{[NONMETRICITY]}\label{NONMETRICITY}\\
R^a_{\phantom{a}b}=0, &&\qquad\mbox{[CURVATURE]}\label{CURVATURE}\\
DM^{ab}-m^{ab}-D\mu^{ab}=\sigma^{ab}, &&\qquad[\mbox{ZEROTH}]\label{ZEROTH}\\
DH_a-E_a=\Sigma_a, &&\qquad\mbox{[FIRST]}\label{FIRST}\\
DH^a_{\phantom{a}b}-E^a_{\phantom{a}b}-\mu^{a}_{\phantom{a}b}-D\nu_b^{\phantom{a}a}=\Delta^a_{\phantom{a}b}, &&\qquad\mbox{[SECOND]}\label{SECOND}\\
\frac{\delta L_{Matt}}{\delta\Phi}=0, \qquad \frac{\delta L_{Matt}}{\delta\Psi^a}=0. &&\qquad\mbox{[MATTER]} \label{MATTER}
\end{eqnarray}

It can be shown that the [ZEROTH] field equation is redundant when all the other equations are true using the Noether identities (\ref{noether-g2}) and (\ref{noether-m2}).  Therefore, the [ZEROTH] equation can be removed without loss of generality.
The [SECOND] field equation, determines part of the Lagrange multipliers, $\mu^{a}_{\phantom{a}b}+D\nu_b^{\phantom{a}a}$
and therefore, we can also remove the [SECOND] field equation without loss of generality.

\subsection{Gauge Choice}
Equation \eqref{CURVATURE} can be solved to determine the spin connection
\begin{equation}
\omega^a_{\phantom{a}b}=(L^{-1})^a_{~c}dL^c_{~b},
\end{equation}
for some invertible matrix $L^a_{~b}(x^\mu) \in GL(4,\mathbb{R})$ which are local functions of the coordinates.  Because we have assumed the {\em{Principle of Relativity}} we have complete freedom to choose a gauge so that computations are simplified. A convenient choice of gauge useful in teleparallel theories of gravity is the orthonormal gauge in which the tangent space metric is diagonalized to become $g_{ab}=\mbox{Diag}[-1,1,1,1]$.  In this gauge, equation \eqref{NONMETRICITY} can be easily solved to yield
\begin{equation}
\omega_{(ab)}=0.
\end{equation}
With this gauge choice, there is still a group of residual gauge transformations possible, that of local orthogonal transformations $O(1,3)$, or any restrictions thereof.
It is also possible to choose a proper orthonormal gauge, in which the metric is diagonalized to   $g_{ab}=\mbox{Diag}[-1,1,1,1]$ and the spin connection becomes $\omega^a_{\phantom{a}b}=0$.  If this gauge choice is made, the group of residual gauge transformations are global (constant) orthogonal transformations $O(1,3)$, or any restrictions thereof.

As it is abundantly clear via the \emph{Principle of Relativity}, we have the freedom to choose to work in an orthonormal frame in which the spin connection may be non-trivial and  we have the freedom to choose to work in a proper orthonormal frame if desired.  Our decision on which frame choice might be informed by the type of problems one wishes to analyze.  In either case we obtain from equation \eqref{conto_LC}
\begin{equation}
\omega^a_{~b}=\omegaLC^a_{~b}+K^a_{~b}.
\end{equation}
Additionally, from equation \eqref{conn_LC}, the Levi-Civita spin connection $\omegaLC^a_{~b}$ in an orthonormal gauge (proper or not)
\begin{equation}\label{42}
\omegaLC_{ab} :=  -e_{[a}\rfloor C_{b]} +\frac{1}{2}(e_a\rfloor e_b\rfloor C_c)\,h^c,
\end{equation}
is a function only of the anholonimity.  By definition of the Levi-Civita connection, the corresponding non-metricity one-form $\QLC_{ab}$ and torsion two-form $\TLC^a$ are identically zero. Only the curvature two-form $\RLC^a_{~b}$ is non-zero.  The zero torsion condition for the Levi-Civita connection yields
\begin{equation}
\TLC^a :=\DLC h^a= dh^a+ \omegaLC^a_{\phantom{a}b} \wedge h^b =C^a+ \omegaLC^a_{\phantom{a}b} \wedge h^b =0,
\end{equation}
which yields the following relation between the anholonimity and the Levi-Civita connection
\begin{equation}\label{def_anholo}
C^a=-\omegaLC^a_{\phantom{a}b}\wedge h^b,
\end{equation}
in any orthonormal gauge. Note this is just the inverse of equation \eqref{42}.

\subsection{Hyper-momentum, spin current and the Noether Identities}

In the case of scalar matter only where $L_{Matt}$ is independent of $\Psi^a$ but continues to depend on the torsion $T^a$, the Noether identity \eqref{noether-m2} yields
\begin{equation}
D\tau_{ab}=-h_{[a} \wedge \Sigma_{b]},
\end{equation}
highlighting the relationship between the spin current (part of the hyper-momentum),
\begin{equation}
\tau_{ab}:=\Delta_{[ab]} =   \frac{1}{2}\left(h_{a}\wedge \frac{\partial L_{Matt}}{\partial T^{b}}-h_{b}\wedge \frac{\partial L_{Matt}}{\partial T^{a}}\right)
\end{equation}
and the antisymmetric part of the canonical energy momentum $\Sigma_a$.
Further, if $L_{Matt}$ is independent of both $\Psi^a$ and $T^a$, then there is no hyper-momentum and consequently the antisymmetric and symmetric parts of equation \eqref{noether-m2} yield
\begin{equation}\label{3.17}
h_{[a} \wedge \Sigma_{b]} =0, \qquad \text{and} \qquad h_{(a} \wedge \Sigma_{b)} =\sigma_{ab},
\end{equation}
showing how the metrical energy momentum, $\sigma_{ab}$ is equivalent to canonical energy momentum $\Sigma_a$ when there is no hyper-momentum source in teleparallel gravity.

Interestingly, in teleparallel gravity the Noether identity associated with \emph{General Covariance} \eqref{noether-m1} becomes
\begin{equation} \label{Dsigma}
D\Sigma_a=(e_a\rfloor T^b)\wedge \Sigma_b.
\end{equation}
If the covariant derivative in equation \eqref{Dsigma} is expressed in terms of the Levi-Civita covariant derivative and the contorsion, then using the symmetry properties of the contorsion we arrive at
\begin{equation}
\DLC \Sigma_a = (e_a\rfloor K^{bc})\,h_{[c}\wedge \Sigma_{b]}.
\end{equation}
Again, if $L_{Matt}$ is independent of $\Psi^a$ and $T^a$, then using equation \eqref{3.17} this equation reduces to the usual Noether identity (energy momentum conservation equation) for GR,
\begin{equation}
\DLC \Sigma_a = 0.
\end{equation}

\subsection{Source Coupling and the Matter Lagrangian}\label{source coupling}

In order to properly incorporate matter in a teleparallel theory of gravity, let us take a closer look at how matter source fields couple to the geometry.  Let $\Phi$ be a scalar-valued source field and let $\Psi^a$ be a vector valued $p$-form source field.

\subsubsection{Minkowski -- Holonomic and Inertial}
In a Minkowski geometry, the basis for special relativistic field theories, let us assume a holonomic co-frame basis $\tilde{h}^a=\tilde{h}^a_{~\mu}\, dx^\mu$ and a trivial spin connection initially. This is by construction an inertial co-frame.  
The geometrical quantities of non-metricity, torsion and curvature are all identically zero. Of particular note, the anholonimity $\tilde{C}^a:=d\tilde{h}^a=0$ indicating how our initial frame choice is holonomic.
The space-time metric is $\tilde{g}^{\vphantom{\mu}}_{\mu\nu}=g^{\vphantom{\mu}}_{ab}\tilde{h}^a_{~\mu}\tilde{h}^b_{~\nu}$ which is a metric in a Minkowski geometry.
In such a framework, a suitable matter Lagrangian density is of the form
\begin{equation}\label{L_matt_mink}
\mathcal{L}_{Matt}:=\mathcal{L}_{Matt}(g_{ab},\tilde{h}^a,\Phi,d\Phi,\Psi^a,d\Psi^a).
\end{equation}

\subsubsection{Minkowski -- non-Holonomic and non-Inertial}
If we change our co-frame and apply a local Lorentz transformation, $\Lambda^a_{~b}$, we find our Lorentz transformed co-frame basis $\hat{h}^{a}=\Lambda^a_{~b}\tilde{h}^b$ becomes non-holonomic since
\begin{equation}
\hat{C}^a:=d\hat{h}^a =d\Lambda^a_{~c}\wedge (\Lambda^{-1})^c_{~b}\,\hat{h}^b \not = 0.
\end{equation}
Additionally, due to the non-homogeneous transformation rules for spin connections under frame transformations, a non-trivial spin connection arises
\begin{equation}
\hat{\omega}^a_{~b}=\Lambda^a_{~c}d(\Lambda^{-1})^c_{~b}=-d\Lambda^a_{~c}(\Lambda^{-1})^c_{~b},
\end{equation}
which encodes the inertial effects of moving to a non-inertial co-frame. The spin connection $\hat{\omega}^a_{~b}$ is called a Lorentz spin connection since it is the result from moving from an inertial co-frame to a non-inertial co-frame via a Lorentz transformation.
The geometrical quantities of non-metricity, torsion and curvature all remain identically zero since a Lorentz transformation does not alter the geometrical properties of Minkowski geometry.  We observe from the definition of the torsion in a Minkowski geometry described by $\hat{h}^a$ and $\hat\omega^a_{~b}$ that
\begin{equation}\label{torsion_in_mink}
\hat{T}^a:=\hat{D}\hat{h}^a=d\hat{h}^a+\hat{\omega}^a_{~b}\wedge \hat{h}^b=0\  \text{and so}\   \hat{C}^a=-\hat{\omega}^a_{~b}\wedge \hat{h}^b,
\end{equation}
clearly showing how the inertial effects of moving to a non-holonomic co-frame manifest themselves in a non-trivial spin connection $\hat{\omega}^a_{~b}$.
Since we have chosen an orthonormal gauge, equation \eqref{torsion_in_mink} yields the relationship
\begin{equation}\label{inv_C}
\hat{\omega}_{ab}=-\left(\hat{e}_{[a}\rfloor \hat{C}_{b]} - \frac{1}{2}(\hat{e}_a\rfloor \hat{e}_b\rfloor \hat{C}_c)\, \hat{h}^c\right),
\end{equation}
where $\hat{e}_a$ are the corresponding dual vectors, i.e., $\hat{e}_a\rfloor \hat{h}^b=\delta^a_{b}$.
Further, since $\DLC \hat{h}^a=0$ by definition, we observe that the Lorentz spin connection and the Levi-Civita spin connection in the absence of gravity are equal
\begin{equation}\label{equiv inert and LC}
\hat{\omega}^a_{~b}=\omegaLC^a_{~b}.
\end{equation}
The space-time metric becomes
\begin{equation}
\hat{g}^{\vphantom{\mu}}_{\mu\nu}:=g^{\vphantom{\mu}}_{ab}\hat{h}^a_{~\mu}\hat{h}^b_{~\nu}
=g^{\vphantom{\mu}}_{ab}\Lambda^a_{~c}\tilde{h}^c_{~\mu}\Lambda^b_{~d}\tilde{h}^d_{~\nu}=\tilde{g}_{\mu\nu},
\end{equation}
and continues to represent the metric in a Minkowski geometry.
The matter Lagrangian density in a special relativistic field theory in a non-inertial co-frame is
\begin{equation}\label{Matt2}
\mathcal{L}_{Matt}=\mathcal{L}_{Matt}(g_{ab},\hat{h}^a,\Phi,d\Phi,\Psi^a,\hat{D}\Psi^a)
\end{equation}
where $\hat{D}$ is now the covariant exterior derivative with respect to the Lorentz spin connection $\hat{\omega}^a_{~b}$\footnote{$\hat{D}\Psi^a$ could be replaced by $\DLC \Psi^a$ in the absence of gravity because $\hat{\omega}^a_{~b}=\omegaLC^a_{~b}$.}.

\subsubsection{The Gauging of Teleparallel Gravity}

Teleparallel gravity can be framed as a gauge theory for the translation group \cite{Aldrovandi_Pereira2013,Krssak:2018ywd}.  If one is initially in an inertial frame $\tilde{h}^a$, gravity manifests itself as a vector-valued one-form field $\tilde{B}^a$.  Its field strength $\tilde{F}^a$, in the inertial frame is simply the exterior derivative
\begin{equation}
\tilde{F}^a:=d\tilde{B}^a.
\end{equation}
Moving to a non-inertial frame via a local Lorentz transformation, $\Lambda^a_{~b}$, the field strength becomes
\begin{equation}\label{FS2}
{F}^a:=\hat{D}{B}^a=d{B}^a+\hat{\omega}^a_{~b}\wedge {B}^b,
\end{equation}
where ${B}^a=\Lambda^a_{~b}\tilde{B}^b$.
The field strength ${F}^a$ of the gauge field ${B}^a$ is just the torsion of the super-position of the inertial frame and the gauge field.
Since $\hat{D}\hat{h}^a=0$, by adding equation \eqref{torsion_in_mink} to equation \eqref{FS2} we find
\begin{equation}\label{FS}
{F}^a=\hat{D}{B}^a + \hat{D}\hat{h}^a = \hat{D}(\hat{h}^a+{B}^a)= \hat{D}h^a=T^a
\end{equation}
which is the torsion $T^a$ of the co-frame constructed from the sum
\begin{equation}
h^b=\hat{h}^b+B^b,
\end{equation}
with respect to the Lorentz connection $\hat{\omega}^a_{~b}$.

\subsubsection{Minimal Coupling Prescription}\label{minimal-coupling}

The final piece of the puzzle requires one to couple the geometry to the matter fields.  To do so requires one to propose a mechanism in which the gravitational fields $B^a$ or $h^a$ couple to the special relativistic matter fields, $\Phi$ and $\Psi^a$.   For example in GR, the typical coupling prescription to transform a special relativistic matter Lagrangian to a general relativistic matter Lagrangian is to apply the mappings
\begin{equation}\label{min_coupling}
\eta_{\mu\nu}\to g_{\mu\nu}  \text{\qquad and\qquad } \partial_\mu \to \nablaLC_\mu \text{\ (or equivalently \ } d \to \DLC)
\end{equation}
into a special relativistic matter Lagrangian.  This is the minimal coupling prescription for GR which ensures that the equivalence principle is satisfied.  The resulting field equations for the matter fields in GR arising from the variation of the general relativistic matter Lagrangian are the same as those computed by applying the coupling principle to the special relativistic field equations \cite{Saa:1996mt,Delhom:2020hkb}.

In non-Riemannian theories of gravity, including teleparallel gravity, the minimal coupling prescription \eqref{min_coupling} does not universally apply in the same way \cite{Saa:1996mt,Delhom:2020hkb}. The simple application of a coupling  prescription may yield differences between the field equations when computed from the transformed matter Lagrangian and the transformed special relativistic field equations after applying the coupling prescription \cite{Saa:1996mt,Delhom:2020hkb}.  Therefore, any coupling prescription in teleparallel gravity must be considered carefully.

\subsubsection{Translational Coupling Prescription}

For teleparallel gravity, a proposal to consistently couple the geometry to the matter fields is the translational coupling prescription \cite{Aldrovandi_Pereira2013}
\begin{equation}\label{trans_coupling}
\hat{h}^a \to h^a=\hat{h}^a+B^a.
\end{equation}
Can  the matter Lagrangian density for special relativistic field theories be adapted in the presence of gravity by simply applying equation \eqref{trans_coupling} in equation \eqref{Matt2}?

We note that the translational coupling prescription \eqref{trans_coupling} introduces a general Riemannian space-time metric $g^{\vphantom{\mu}}_{\mu\nu}=g^{\vphantom{\mu}}_{ab}h^a_{~\mu}h^b_{~\nu}$, along the same lines as one finds in the minimal coupling principle for GR described in equation \eqref{min_coupling}.
However, in equation \eqref{Matt2}, in addition to $\hat{h}^a$ there exists the derivative term $\hat{D}\Psi^a$.  How does the covariant derivative $\hat{D}\Psi^a$ transform? In particular, how does the Lorentz connection $\hat{\omega}^a_{~b}$ transform under the translational coupling prescription \eqref{trans_coupling}?

Before we complete the coupling prescription to handle derivatives, we will need the inverse of the torsion $T^a=\hat{D}h^a$, when expressed in terms of the contorsion. From equations \eqref{def_conto} and \eqref{FS} we have
\begin{equation}
T^a=\hat{D}h^a=K^a_{~b}h^b = C^a+\hat{\omega}^a_{~b}\wedge h^b
\end{equation}
where $C^a=dh^a$.  Given that $\hat{\omega}_{(ab)} = 0$, this expression can be inverted to yield
\begin{equation}\label{Komega_inverse}
\hat{\omega}_{ab}-K_{ab}=-{e}_{[a}\rfloor {C}_{b]}
 + \frac{1}{2}\left({e}_a\rfloor {e}_b\rfloor {C}_c\right){h}^c.
\end{equation}

Therefore under the translational coupling prescription $\hat{h}^a\to h^a$, $\hat{e}_a \to e_a$ and $\hat{C}^a \to C^a$ the Lorentz connection $\hat{\omega}^a_{~b}$ defined in equation \eqref{inv_C} transforms as
\begin{align}
\hat{\omega}_{ab}
&\to -{e}_{[a}\rfloor {C}_{b]} + \frac{1}{2}({e}_a\rfloor {e}_b\rfloor {C}_c)\, {h}^c ,
\end{align}
and using equations \eqref{Komega_inverse} and \eqref{conto_LC}, with $Q_{ab}=0$ we find that
\begin{align}
\hat{\omega}_{ab} &\to \hat{\omega}_{ab}-K_{ab} = \omegaLC_{ab}. \qquad\qquad\label{62}
\end{align}
The translational coupling principle $\hat{h}^a \to h^a=\hat{h}^a+B^a$ imposes a corresponding mapping of the covariant derivative  $\hat{D} \to \DLC$ in the matter Lagrangian density.   We also note that equation \eqref{62} shows how the Levi-Civita connection in the presence of gravity contains two terms, one is the result of inertial effects, and one because of the existence of gravity. We therefore conclude that the translational coupling prescription \eqref{trans_coupling} when applied to the special relativistic matter Lagrangian in equation \eqref{Matt2} yields a matter Lagrangian density of the form (see also derivations and more physical arguments in \cite{Aldrovandi_Pereira2013})
\begin{equation}\label{min_coup_L}
\mathcal{L}_{Matt}:=\mathcal{L}_{Matt}(g_{ab},h^a,\Phi,d\Phi,\Psi^a,\DLC\Psi^a),
\end{equation}
suitable for use in teleparallel theories of gravity.

The matter Lagrangian density \eqref{min_coup_L} and its associated field equations $\frac{\delta L_{Matt}}{\delta \Phi}$ and $\frac{\delta L_{Matt}}{\delta \Psi^a}$ are independent of the torsion. In this way, we say that the translational coupling prescription results in a \emph{minimally coupled} gravitational theory.
Of course the gravitational fields can be non-minimally coupled to the source fields by introducing the torsion non-trivially into the matter Lagrangian density or by assuming $D\Psi^a$ instead of $\DLC\Psi^a$.

\subsection{Example: \texorpdfstring{$f(T)$}{f(T)}  teleparallel gravity with scalar matter}

On a four dimensional differentiable manifold with metric, the torsion 2-form has three irreducible parts invariant under the general linear group \cite{Hehl_McCrea_Mielke_Neeman1995}, called TENTOR ${}^{(1)}T^a$, TRATOR ${}^{(2)}T^a$, and AXITOR ${}^{(3)}T^a$, having $16$, $4$ and $4$ independent components respectively
\begin{equation}
T^a={}^{(1)}T^a+{}^{(2)}T^a+{}^{(3)}T^a.
\end{equation}
For a metric having Lorentzian signature the irreducible parts of the torsion are given by
\begin{subequations}
\begin{eqnarray}
{}^{(2)}T^a&:=&\frac{1}{3}h^a\wedge h_b\rfloor T^b,  \\
{}^{(3)}T^a&:=&-\frac{1}{3}{}^*\!\left(h^a\wedge P\right), \quad\mathrm{with}\quad P={}^*\!\left(T^a\wedge h_a\right),\\
{}^{(1)}T^a&:=&T^a-^{(2)}T^a-^{(3)}T^a.
\end{eqnarray}
\end{subequations}
A special torsion scalar $T$ can constructed from the following quadratic scalars
\begin{eqnarray*}
T_{TEN}&:=&{}^*\left({}^{(1)}T^a\wedge{}^*{}^{(1)}T_a\right),\\
T_{TRA}&:=&{}^*\left({}^{(2)}T^a\wedge{}^*{}^{(2)}T_a\right),\\
T_{AXI}&:=&{}^*\left({}^{(3)}T^a\wedge{}^*{}^{(3)}T_a\right).
\end{eqnarray*}
Summing the above quadratic scalars with the following coefficients \cite{Hehl_McCrea_Mielke_Neeman1995}
\begin{subequations}
\begin{eqnarray}
T &:=& -T_{TEN}+2T_{TRA}+\frac{1}{2} T_{AXI},\\
  &=&{}^*(T^a\wedge {}^*S_a),
\end{eqnarray}
\end{subequations}
yields the torsion scalar $T$.  The super-potential is defined to be
\begin{subequations}
\begin{eqnarray}
{}^*S_a&:=& -{}^*{}^{(1)}T_a +2{}^*{}^{(2)}T_a+\frac{1}{2}{}^*{}^{(3)}T_a,\\
&=&-\frac{1}{2}K^{cd}\wedge \eta_{acd},
\end{eqnarray}
\end{subequations}
which we observe is simply related to the dual of the contorsion.

An example of a teleparallel theory of gravity is TEGR \cite{Aldrovandi_Pereira2013} which is dynamically equivalent to GR and is based on the scalar $T$.  The gauge potential 4-form for TEGR is
\begin{equation}
V_{gauge}:=-\frac{1}{2\kappa^2}T\,\eta,
\end{equation}
where $\kappa^2$ is the coupling constant. The equivalence of TEGR and GR is the result of the identity
\begin{equation}
\RLC \,\eta=-T\,\eta+2d(h^a\wedge {}^*T_a)
\end{equation}
where $\RLC$ is the usual Ricci scalar computed from the metric and the term $2d(h^a\wedge {}^*T_a)$ yields a surface term when placed in the action.

A popular generalization of TEGR is the class of $f(T)$ teleparallel gravity theories \cite{Krssak_Saridakis2015,Krssak:2018ywd}. New General Relativity \cite{Hayashi:1979qx} and other teleparallel theories of gravity, $f(T,B)$, etc., are also possible (see \cite{Bahamonde:2021gfp} for a recent comprehensive review of teleparallel gravity in general). As an example, we shall show the field equations for $f(T)$ teleparallel gravity.
For the class of $f(T)$ teleparallel gravity theories, the gauge potential depends on a twice differentiable function $f$ of the scalar torsion $T$. The gauge potential 4-form is
\begin{equation}
V_{gauge}:=-\frac{1}{2\kappa^2}f(T)\,\eta.\nonumber
\end{equation}

With the gauge potential $V_{gauge}$ defined,  we compute
\begin{eqnarray}
H_a&=&-\frac{1}{\kappa^2}f'(T)\,{}^*S_a, \\
E_a &=&     -\frac{1}{2\kappa^2}f(T) e_a\rfloor\eta - \frac{1}{\kappa^2} f'(T) e_a\rfloor T^b \wedge {}^*S_b.\label{def_E}
\end{eqnarray}
The remaining field equations are [FIRST] and the [MATTER], equations \eqref{FIRST} and \eqref{MATTER}. Equation \eqref{FIRST} can be decomposed into a symmetric and antisymmetric parts. The complete set of remaining field equations given that the matter source is a scalar matter source are
\begin{eqnarray*}
\kappa^2 \Sigma_{(a} \wedge h_{b)}   &=& -f''(T)dT\wedge {}^*S_{(a}\wedge h_{b)}-f'(T)\Bigl(D{}^*S_{(a}-e_{(a}\rfloor T^c\wedge {}^*S_{\underline{c}}\Bigr)\wedge h_{b)}+\frac{1}{2}f(T)g_{ab}\eta ,\\
\kappa^2 \Sigma_{[a} \wedge h_{b]}   &=&  -f''(T)\,dT\wedge {}^*S_{[a}\wedge h_{b]} ,\\
0&=&\frac{\partial L_{Matt}}{\partial\Phi}-d\left(\frac{\partial L_{Matt}}{\partial d\Phi}\right).\\
\end{eqnarray*}

It should be mentioned that the class of $f(T)$ teleparallel gravity theories continues to be investigated for its viability as a robust theory of gravity \cite{Golovnev:2020zpv,Golovnev:2024sod}. There is no agreement on the number of physical degrees of freedom in the theory  \cite{Li:2011rn,Ferraro:2018axk,Ong:2018heg,Blixt:2020ekl,Blagojevic:2020dyq}. Indeed, the number of degrees of freedom appears to be dependent on the background geometry \cite{Chen:2014qtl}. Further, perturbative approaches to $f(T)$ teleparallel gravity indicate a strong coupling problem \cite{BeltranJimenez:2020fvy,Hu:2023juh} which would introduce a ghost into the theory \cite{Deffayet:2005ys}. All this to say $f(T)$ teleparallel gravity has not yet been proven to be a viable theory, and therefore still requires a thorough analysis. However, $f(T)$ teleparallel gravity does provide an interesting arena to play with and test new ideas.


\section{Spherically Symmetric geometries in Teleparallel Gravity}

\subsection{Affine Frame Symmetries}

Choosing coordinates $ x^\mu = \{t, r, \theta, \phi\}$ adapted to the symmetry of the situation, the three vector generators ${\bf X}_{(I)}$ for $I=1,2,3$ determining spherical symmetric geometries are,
\begin{subequations} \label{vectorgenerators}
\begin{align}
{\bf X}_{(1)} &:= \hphantom{-}\,\partial_{\phi},\\
{\bf X}_{(2)} &:= - \cos \phi \,\partial_{\theta} + \frac{\sin \phi}{\tan \theta} \,\partial_{\phi},\\
{\bf X}_{(3)} &:= \hphantom{-}\sin \phi \,\partial_{\theta} + \frac{\cos \phi}{\tan \theta} \,\partial_{\phi}.
\end{align}
\end{subequations}
Affine frame symmetries generated by vectors ${\bf X}_{(I)}$ can be defined through their action on the Lie derivative of the co-frame and connection
\begin{subequations}
\begin{align}
{\cal{L}}_{{\bf X}_{(I)}} h^a &= \lambda^{~~a}_{(I)b}h^b, \label{affine1}\\
({\cal{L}}_{{\bf X}_{(I)}} {\bm{\omega}})^a_{~bc} &= 0,\label{affine2}
\end{align}
\end{subequations}
where $\lambda^{~~a}_{(I)\,b}$ is the Lie algebra generator for some element of the isotropy group associated with the affine frame symmetry \cite{Hohmann:2019nat,McNutt:2023nxm}.  If there is no isotropy group, then $\lambda^{~~a}_{(I)b}=0$. Further, equation \eqref{affine2} can be equivalently expressed in terms of the Lie derivative of the spin connection one-forms and the negative exterior covariant derivative of these same $\lambda^{~~a}_{(I)\,b}$,
\begin{align}
{\cal{L}}_{{\bf X}_{(I)}} \omega^a_{~b} &= -D\lambda^{~~a}_{(I)b},
\end{align}
clearly showing how the existence of an isotropy complicates the determination of both the frame and spin connection components.

Since spherically symmetric geometries contain an isotropy group, determining the co-frame and spin connection associated with teleparallel spherically symmetric geometries is complicated by these additional quantities $\lambda^{~~a}_{(I)\,b}$ which are local functions of the coordinates. A procedure to find a canonical basis for the co-frame $h^a$ and its corresponding spin connection $\omega^a_{~b}$ respecting the affine frame symmetries dictated by \eqref{vectorgenerators} was developed in \cite{McNutt:2023nxm} using the Cartan-Karlhede algorithm to fix the additional quantities $\lambda^{~~a}_{(I)\,b}$ as much as possible in an invariantly defined way.

\subsection{The Diagonal Frame}\label{section-diagonal-frame}

It has been determined that an appropriate orthonormal frame ansatz  $h^a$ and its corresponding spin connection $\omega^a_{~b}$ for a general spherically symmetric teleparallel geometry is of the form \cite{McNutt:2023nxm}
\begin{subequations}\label{diagonal_frame}
\begin{align}
h^1&=A_1(t,r)\,dt, \\
h^2&=A_2(t,r)\,dr,\\
h^3&=A_3(t,r)\,d\theta,\\
h^4&=A_3(t,r)\sin\theta\, d\phi,
\end{align}
\end{subequations}
with
\begin{subequations} \label{orthonormal_connection}
\begin{align}
\omega^1_{~2} &= \hphantom{-} \Big(\partial_{t} \psi(t, r)\Big) \,dt + \Big(\partial_{r} \psi(t, r)\Big) \,dr, \\
\omega^1_{~3} &= -\Big(\sinh{\psi(t, r)} \cos{\chi(t, r)}\Big)\,d\theta -\Big(\sinh{\psi(t, r)} \sin{\chi(t, r)}\sin\theta\Big)\,d\phi, \\
\omega^1_{~4} &= -\Big(\sinh{\psi(t, r)} \sin{\chi(t, r)}\Big)\,d\theta -(\sinh{\psi(t, r)} \cos{\chi(t, r)}\sin\theta)\,d\phi, \\
\omega^2_{~3} &= \hphantom{-}  \Big(\cosh{\psi(t, r)} \cos{\chi(t, r)}\Big)\,d\theta +\Big(\cosh{\psi(t, r)} \sin{\chi(t, r)}\sin\theta\Big)\,d\phi, \\
\omega^2_{~4} &= -\Big(\cosh{\psi(t, r)} \sin{\chi(t, r)}\Big)\,d\theta +\Big(\cosh{\psi(t, r)} \cos{\chi(t, r)}\sin\theta\Big)\,d\phi, \\
\omega^3_{~4}  &=-\Big(\partial_{t} \chi(t, r)\Big) \,dt - \Big(\partial_{r} \chi(t, r)\Big)\,dr - \Big(\cos{\theta}\Big)\,d\phi,
\end{align}
\end{subequations}
\noindent
where $A_1(t,r), A_2(t,r), A_3(t,r), \psi(t,r)$, and $\chi(t,r)$ are five arbitrary functions. We label such a frame choice a ``diagonal'' orthonormal frame, since the veirbein matrix with our coordinate choice is of diagonal form
\begin{equation}
h^a_{\ \mu} =
\begin{bmatrix}
A_1(t, r) & 0 & 0 & 0 \\
0 & A_2(t, r) & 0 & 0 \\
0 & 0 & A_3(t, r) & 0 \\
0 & 0 & 0 & A_3(t, r) \sin(\theta)
\end{bmatrix}.
\end{equation}
If we insist on using a diagonal frame then there are five arbitrary functions required to describe the most general spherically symmetric teleparallel geometry, $\{A_1(t,r), A_2(t,r), A_3(t,r), \psi(t,r),\chi(t,r)\}$.  This frame and connection choice is not a proper frame since the spin connection is non-trivial.


\subsection{Coordinate Choices}\label{Coordinates}

As we have complete freedom to choose new coordinates, any coordinate transformation of the form  $t=\alpha(t',r')$ and $r=\beta(t',r')$ could be chosen.  We observe that $\omega^1_{~2}=d\psi(t,r)$ and $\omega^3_{~4}=-d\chi(t,r)-\cos(\theta)\,d\phi$, and since $\psi(t,r)$ and $\chi(t,r)$ are scalars, the nature of the spin connection \eqref{orthonormal_connection} does not change under this coordinate transformation.  However,  the diagonal frame \eqref{diagonal_frame} under this coordinate transformation becomes
\begin{subequations}\label{non-diagonal_frame}
\begin{align}
h^1&=A_1(\alpha(t',r'),\beta(t',r'))\partial_{t'}(\alpha(t',r'))\,dt' + A_1(\alpha(t',r'),\beta(t',r'))\partial_{r'}(\alpha(t',r'))\,dr', \\
h^2&=A_2(\alpha(t',r'),\beta(t',r'))\partial_{t'}(\beta(t',r'))\,dt'+ A_2(\alpha(t',r'),\beta(t',r'))\partial_{r'}(\beta(t',r'))\,dr',\\
h^3&=A_3(\alpha(t',r'),\beta(t',r'))\,d\theta,\\
h^4&=A_3(\alpha(t',r'),\beta(t',r'))\sin\theta\, d\phi,
\end{align}
\end{subequations}
and will become non-diagonal except in some special cases.  If $\partial_{r'}\alpha=\partial_{t'}\beta=0$, then the co-frame in these new coordinates is diagonal.  For example, in the static spherically symmetric case we can choose $\alpha(t',r')=t'$ and $\beta(t',r')=\beta(r')$ and our frame \eqref{non-diagonal_frame} in the new coordinates $t',r'$  is diagonal.  Similarly, in the Kantowski-Sachs case we can choose $\alpha(t',r')=\alpha(t')$ and $\beta(t',r')=r'$ and our frame \eqref{non-diagonal_frame} in the new coordinates is once again diagonal.

There is a second case we mention only for completeness.  If $\partial_{t'}\alpha=\partial_{r'}\beta=0$, then the resulting co-frame can be made diagonal by simply applying an orthogonal transformation that swaps $h^1$ and $h^2$.  However, this results in a situation in which the $t'$ coordinate no longer represents time.   The orthogonal transformation that swaps $h^1$ and $h^2$ is not a proper ortho-chronous Lorentz transformation.  We do not consider this situation further.

If we are willing to drop the diagonal requirement on the co-frame ansatz, then there is an opportunity to use the coordinate freedom to reduce the number of arbitrary functions from five to four. We break the choices of coordinates down into a number of cases.  In each case we can show that there are four arbitrary functions remaining and in all cases except the most general case, the diagonal form of the co-frame is maintained.

\paragraph{$A_3$ is constant:}
When $A_3(t,r)=a_3$, a constant, then the number of arbitrary functions describing this spherically symmetric teleparallel geometry is four. New coordinates are not required, The resultant geometry can be considered as a Cartesian product of a 2-sphere and 2-dimensional surface. The co-frame maintains its diagonal form.

\paragraph{$A_3$ is not constant, Kantowski-Sachs Case:}
In the Kantowski-Sachs case all the arbitrary functions become independent of the coordinate $r$.  We can choose new coordinates of the form $t'=\int A_1(t)\, dt$ and $r'=r$. The Kantowski-Sachs frame maintains its diagonal form and
the frame connection pair now contains only four arbitrary functions of $t'$, $\{A_2(t'),A_3(t'),\psi(t'),\chi(t')\}$. Alternatively, one could also choose new coordinates of the form $t'=A_3(t)$ and $r'=r$ which again reduces the number of arbitrary functions down to four and maintains the diagonal form of the co-frame.  However, this coordinate choice for the Kantowski-Sachs geometry is not popular in the literature but is analogous to what is done in the static and general case below.

\paragraph{$A_3$ is not constant, Static Case:}
In the static case all the arbitrary functions become independent of the coordinate $t$.  We can choose new coordinates of the form $t'=t$ and $r'=A_3(r)$. The static frame in these ``curvature coordinates'' maintains its diagonal form and the frame connection pair now contains only four arbitrary functions of $r'$, $\{A_1(r'),A_2(r'),\psi(r'),\chi(r')\}$.  Alternatively, ``isotropic coordinates'' can be chosen with the coordinate choice $t'=t$ and $r'=A_3(r)/A_2(r)$ again the number of functions reduces to four and the co-frame maintains its diagonal form.

\paragraph{$A_3$ is not constant, General Case:}
If $A_3(t,r)$ is not a constant, then new coordinates can always be chosen so that the number of arbitrary functions becomes four.  If $\partial_r A_3(t,r) \not=0$ then we can choose new coordinates of the form $t'=t$ and $r'=A_3(t,r)$. The resulting co-frame connection pair has only four arbitrary functions, but the co-frame is no longer diagonal.  If $\partial_r A_3(t,r)=0$ then we can choose new coordinates of the form $r'=r$ and $t'=A_3(t,r)$ resulting in a non-diagonal co-frame connection pair containing four arbitrary functions. Again, a coordinate choice $t'=t$ and $r'=A_3(t,r)/A_2(t,r)$ reduces the number of functions to four but the resulting co-frame is not diagonal.

Of course there are an infinite number of coordinate choices, some of which may lend themselves better to different physical scenarios.  The assumption of a diagonal frame in general is coordinate dependent.

\subsection{The Proper Co-frame}

To determine the proper co-frame ansatz $\ph^a$   for general spherically symmetric teleparallel geometries requires one to find the Lorentz transformation that transforms the spin connection \eqref{orthonormal_connection} to a trivial spin connection.  Any spin connection $\omega^a_{~b}$ in a teleparallel geometry is generated by some local Lorentz transformation, $\Lambda^a_{~b}$ as
\begin{equation} \label{DE_conn}
\omega^a_{~b} = (\Lambda^{-1})^a_{~c}  d \Lambda^c_{~b}.
\end{equation}
If one was to apply the Lorentz transformation $\Lambda^a_{~b}$ to the connection given by \eqref{DE_conn}, the resulting connection is  trivial.  Since, $\omega^a_{~b}$ is given by \eqref{orthonormal_connection} for a general spherically symmetric teleparallel geometry, we wish to determine the corresponding $\Lambda^a_{~b}$ which will then allow us to determine the proper orthonormal frame.  Rearranging equation \eqref{DE_conn} we find
\begin{equation}
d \Lambda^a_{~b} = \Lambda^a_{~c} \omega^c_{~b}.\label{DE_for_lambda}
\end{equation}
Equation \eqref{DE_for_lambda} yields a system of linear partial differential equations that can be solved for the components, $\Lambda^a_{~b}$.  The integrability condition of \eqref{DE_for_lambda} is the zero curvature requirement of the connection $\omega^a_{~b}$.

Writing the matrix $\Lambda = \Lambda^a_{~b}$ and similar for other matrices a solution to \eqref{DE_for_lambda} can be determined. The Lorentz transformation that generates the most general spherically symmetric spin connection includes a constant Lorentz transformation, $C$ multiplied by three rotations and one boost in the following construction,
\begin{equation}
      \Lambda = C \cdot R_1(\phi) \cdot R_3(\theta) \cdot R_1(\chi) \cdot B_1(\psi).
\end{equation}
Where
\begin{subequations}
\begin{align}
    R_1(\phi) &= \begin{bmatrix}
        1 & 0 & 0 & 0 \\
0 & 1 & 0 & 0 \\
0 & 0 & \cos(\phi) & -\sin(\phi) \\
0 & 0 & \sin(\phi) & \cos(\phi)
    \end{bmatrix}
   &
     R_3(\theta) &= \begin{bmatrix}
        1 & 0 & 0 & 0 \\
0 & \cos(\theta) & \sin(\theta) & 0 \\
0 & -\sin(\theta) & \cos(\theta) & 0 \\
0 & 0 & 0 & 1
    \end{bmatrix}\\
    R_1(\chi) &= \begin{bmatrix}
        1 & 0 & 0 & 0 \\
0 & 1 & 0 & 0 \\
0 & 0 & \cos(\chi) & -\sin(\chi) \\
0 & 0 & \sin(\chi) & \cos(\chi)
\end{bmatrix}
&
    B_1(\psi) &= \begin{bmatrix}
        \cosh(\psi) & \sinh(\psi) & 0 & 0 \\
\sinh(\psi) & \cosh(\psi) & 0 & 0 \\
0 & 0 & 1 & 0 \\
0 & 0 & 0 & 1
\end{bmatrix}
\end{align}
\end{subequations}
The functions appearing in the spin connection, $\psi(t,r)$ and $\chi(t,r)$ are arbitrary but may be chosen to solve constraints arising from field equations or symmetry requirements.   For example, we can choose $\psi(t,r)$ and $\chi(t,r)$ so that the anti-symmetric part of the covariant $f(T)$ field equations (or the antisymmetric part of the field equations in other teleparallel theories of gravity) are satisfied.

Since the remaining gauge freedom in an orthonormal proper frame are constant (global) Lorentz transformations, we have complete freedom to set the constant Lorentz transformation $C$ to the identity matrix without loss of generality.  The general solution is
\begin{equation}\label{Lorentz-Matrix}
    \Lambda^a_{~b}=\scalemath{0.75}{\begin{bmatrix}
    \cosh(\psi) & \sinh(\psi) & 0 & 0 \\[0.25cm]
    \cos(\theta) \sinh(\psi) & \cosh(\psi) \cos(\theta) & \cos(\chi) \sin(\theta) & -\sin(\theta) \sin(\chi) \\[0.25cm]
    -\sin(\theta) \cos(\phi) \sinh(\psi) & -\cos(\phi) \sin(\theta) \cosh(\psi) & \cos(\theta) \cos(\phi) \cos(\chi) - \sin(\phi) \sin(\chi) & -\cos(\phi) \cos(\theta) \sin(\chi) - \sin(\phi) \cos(\chi) \\[0.25cm]
    -\sin(\phi) \sin(\theta) \sinh(\psi) & -\sin(\phi) \sin(\theta) \cosh(\psi) & \cos(\theta) \sin(\phi) \cos(\chi) + \cos(\phi) \sin(\chi) & -\cos(\theta) \sin(\phi) \sin(\chi) + \cos(\phi) \cos(\chi)
\end{bmatrix}}.
\end{equation}

If we wish to explicitly construct the proper spherically symmetric (non-diagonal) orthonormal frame from the diagonal frame \eqref{diagonal_frame}, we  compute
\begin{equation}
    \ph^{a}=\Lambda^{a}_{~b}h^{b},
\end{equation}
which yields,
\begin{subequations} \label{SS_proper_frame}
\begin{align}
\ph^{1}&= \Big( A_1(t, r) \cosh(\psi(t,r))\Big)\, dt
      + \Big( A_2(t, r) \sinh(\psi(t,r))\Big)\, dr, \\
\ph^{2}& =\Big(A_1(t, r) \sinh(\psi(t,r)) \cos(\theta)\Big)\, dt
      + \Big(A_2(t, r) \cosh(\psi(t,r)) \cos(\theta)\Big)\, dr \nonumber\\
      &\qquad\qquad+ \Big(A_3(t, r) \cos(\chi(t,r)) \sin(\theta)\Big)\, d\theta
                   - \Big(A_3(t, r) \sin(\chi(t,r)) \sin^2(\theta) \Big)\, d\phi, \\
\ph^{3}&= - \Big(A_1(t, r)\sinh(\psi(t,r)) \cos(\phi) \sin(\theta)\Big)\,dt
        - \Big(A_2(t, r)\cosh(\psi(t,r)) \cos(\phi) \sin(\theta) \Big)\,dr \nonumber \\
      &\qquad\qquad + \Big(A_3(t, r) \left(\cos(\theta) \cos(\phi) \cos(\chi(t,r)) - \sin(\phi) \sin(\chi(t,r))\right)\Big)\,d\theta \nonumber\\
      &\qquad\qquad\qquad -\Big(A_{3}(t,r)\sin(\theta)\left(\cos(\phi) \cos(\theta) \sin(\chi(t,r)) + \sin(\phi) \cos(\chi(t,r))\right)\Big)\,d\phi,  \\
\ph^{4}&= - \Big(A_1(t, r) \sinh(\psi(t, r))\sin(\phi) \sin(\theta)\Big)\,dt
        - \Big(A_2(t, r) \cosh(\psi(t, r))\sin(\phi) \sin(\theta)\Big)\,dr \nonumber \\
    &\qquad\qquad +  \Big(A_3(t, r) \left(\cos(\theta) \sin(\phi) \cos(\chi(t,r)) + \cos(\phi) \sin(\chi(t,r))\right)\Big)\,d\theta \nonumber\\
     &\qquad\qquad\qquad + \Big(A_3(t, r)\sin(\theta) \left(-\cos(\theta) \sin(\phi) \sin(\chi(t,r)) + \cos(\phi) \cos(\chi(t,r))\right)\Big)\, d\phi,
\end{align}
\end{subequations}
which again contains the five arbitrary functions $\{A_1(t,r), A_2(t,r), A_3(t,r), \psi(t,r),\chi(t,r)\}$ one of which can be eliminated via an appropriate coordinate transformation (See Section \ref{Coordinates}).  If  one is interested in investigating spherically symmetric teleparallel geometries, one can use the diagonal co-frame and corresponding spin-connection ansatz, equations \eqref{diagonal_frame} and \eqref{orthonormal_connection} or choose the proper spherically symmetric (non-diagonal) co-frame given by equation \eqref{SS_proper_frame} and a trivial spin connection.

\subsection{Another approach}\label{Compare-Hohmann}

Another approach to determine the appropriate ansatz to be used in the study of teleparallel spherically symmetric geometries has been completed by Hohmann et al. in \cite{Hohmann:2019nat}.  The proper orthonormal frame $\pbh^a=\pbh^a_{~\mu}\,dx^\mu$ of Hohmann et al. \cite{Hohmann:2019nat} is described by the veirbein matrix
\begin{equation} \label{hohman_frame}
   \pbh^{a}_{\ \mu}=\scalemath{0.80}{\begin{bmatrix}
         C_1 & C_2 & 0 & 0  \\[0.25cm]
         C_3 \sin(\theta) \cos(\phi) & C_4 \cos(\phi) \sin(\theta) & C_5 \cos(\theta) \cos(\phi) - C_6 \sin(\phi) & -\sin(\theta) \left(C_5 \sin(\phi)+ C_6 \cos(\theta) \cos(\phi)\right)  \\[0.25cm]
         C_3 \sin(\phi) \sin(\theta) & C_4 \sin(\phi) \sin(\theta) & C_5 \cos(\theta) \sin(\phi) + C_6 \cos(\phi)  &\sin(\theta) \left(C_5 \cos(\phi) -C_6 \cos(\theta) \sin(\phi)\right)  \\[0.25cm]
         C_3 \cos(\theta) & C_4 \cos(\theta) & -C_5 \sin(\theta) & C_6\sin^2(\theta)
\end{bmatrix}},
\end{equation}
where the six functions $C_i = C_i(t, r)$, $i = 1, \ldots, 6$  depend on the time and radial coordinates.  Hohmann et al. \cite{Hohmann:2019nat} call this orthonormal proper frame the Weitzenb\"{o}ck gauge.
The corresponding metric is
\begin{equation} \label{metric_hohmann}
\mathbf{g} = -(C_1^2 - C_3^2) dt \otimes dt + 2(C_3C_4 - C_1C_2) dt \otimes dr + (C_4^2 - C_2^2) dr \otimes dr + (C_5^2 + C_6^2) d\Omega \otimes d\Omega
\end{equation}
where $d\Omega \otimes d\Omega = d\theta \otimes d\theta +\sin^2(\theta)d\phi \otimes d\phi $. New coordinates can be chosen so that  $C_1C_2-C_3C_4=0$, which reduces the number of arbitrary functions to five \cite{Hohmann:2019nat}.

The proper spherically symmetric co-frame ansatz, $\pbh^a$ determined in \cite{Hohmann:2019nat} as expressed in equation \ref{hohman_frame} is equivalent to the co-frame $\ph^a$ given in equation \eqref{SS_proper_frame}, up to a constant matrix factor. Comparing the line element for the proper spherically symmetric frame \eqref{SS_proper_frame},
\begin{equation}\label{metric_proper_diagonal}
\mathbf{g} = -A^{2}_{1}(t,r) dt \otimes dt + A^{2}_{2}(t,r) dr \otimes dr + A^{2}_{3}(t,r) d\Omega \otimes d\Omega,
\end{equation}
to equation \eqref{metric_hohmann}, it is possible to determine the $C_i(t,r)$ in terms of our five arbitrary functions $A_1(t,r), \ A_2(t,r), \ A_3(t,r),\  \psi(t,r), \ \chi(t,r)$  so that the two metrics are the same.
Comparing the metric components we find,
\begin{subequations}\label{solution1}
\begin{align}
C_1(t,r)&=\hphantom{-}A_{1}(t,r)\cosh(\psi(t,r)), &   C_2(t,r)&=\hphantom{-}A_2(t,r)\sinh(\psi(t,r)),\\
C_3(t,r)&=\hphantom{-}A_1(t,r)\sinh(\psi(t,r)),  &    C_4(t,r)&=\hphantom{-}A_2(t,r)\cosh(\psi(t,r)),\\
C_5(t,r)&=-A_3(t,r)\cos(\chi(t,r)),   &   C_6(t,r)&=-A_3(t,r)\sin(\chi(t,r)).
\end{align}
\end{subequations}
The metrics \eqref{metric_hohmann} and \eqref{metric_proper_diagonal} are now equal however the co-frames are slightly different.
Recall we can multiply by any constant (global) Lorentz transformation, $C=C^a_{~b}$, without any loss of generality.
So,  multiplying by the matrix
\begin{equation}
C^{a}_{\ b}=\begin{bmatrix}
    1 & 0 & 0 & 0 \\
    0 & 0 & 0 & 1 \\
    0 & -1 & 0 & 0 \\
    0 & 0 & -1 & 0 \\
\end{bmatrix}=R_3(\frac{\pi}{2})\cdot R_1(-\frac{\pi}{2}),
\end{equation}
we arrive at the expression
\begin{equation}
    \ph^{a}=C^{a}_{~b}\pbh^{b}
\end{equation}
where the arbitrary functions are inter-related via equation \eqref{solution1}. Therefore the proper orthonormal co-frame determined by Hohmann et al. \cite{Hohmann:2019nat} is equivalent to the proper orthonormal co-frame in equation \eqref{SS_proper_frame} which has been computed from the diagonal spherically symmetric co-frame and corresponding spin connection presented in \cite{McNutt:2023nxm}.


\section{The Field Equations for \texorpdfstring{$f(T)$}{f(T)} Teleparallel Gravity for Spherically Symmetric Geometries}\label{section5}

The geometry will be assumed to be invariant under the symmetry group generated by the affine frame symmetry vectors \eqref{vectorgenerators}, i.e, the geometry is spherically symmetric.  For the matter content, we shall simply assume a co-moving fluid having energy density $\rho(t,r)$ and anisotropic pressures $P_{r}(t,r)$ and $P_{t}(t,r)$ representing the radial and transverse pressures respectively. The canonical energy momentum for such a setup reduces to
\begin{equation}
\Sigma_a= \begin{bmatrix}
-\rho(t,r)\,\eta_1 \\
P_r(t,r)\,\eta_2 \\
P_t(t,r)\,\eta_3 \\
P_t(t,r)\,\eta_4
\end{bmatrix}.
\end{equation}

\subsection{Torsion Scalar and the Field Equations in the Diagonal Orthonormal Frame}

With the understanding that we still have some coordinate freedom to potentially eliminate one of the frame functions, we express the field equations for $f(T)$ teleparallel gravity in the spherically symmetric orthonormal diagonal gauge.  Using the co-frame given by equation \eqref{diagonal_frame} and spin connection given by equation \eqref{orthonormal_connection} (and dropping the explicit variable dependence) the torsion scalar is
\begin{align}
T=&-4\left(\frac{\partial_t \chi}{A_1}\right)\left(\frac{\sin(\chi)\sinh(\psi)}{A_3}\right)
    +4\left(\frac{\partial_r \chi}{A_2}\right)\left(\frac{\sin(\chi)\cosh(\psi)}{A_3}\right)
    +4\left(\frac{\partial_t \psi}{A_1}\right)\Bigg(\frac{\cos(\chi)\cosh(\psi)}{A_3}
     \nonumber\\
  &
    +\frac{\partial_r A_3}{A_2A_3}\Bigg)-4\left(\frac{\partial_r \psi}{A_2}\right)\left(\frac{\cos(\chi)\sinh(\psi)}{A_3}+\frac{\partial_t A_3}{A_1A_3}\right)
    -4\left(\frac{\partial_r A_1}{A_2 A_1}+\frac{\partial_r A_3}{A_2 A_3}\right)\left(\frac{\cos(\chi)\cosh(\psi)}{A_3}\right)
    \nonumber\\
  &
    +4\left(\frac{\partial_t A_2}{A_1 A_2}+\frac{\partial_t A_3}{A_1 A_3}\right)\left(\frac{\cos(\chi)\sinh(\psi)}{A_3}\right)
  -2\left(\frac{\partial_r A_3}{A_2 A_3}\right)^2
  +2\left(\frac{\partial_t A_3}{A_1 A_3}\right)^2
  -4\left(\frac{\partial_r A_1}{A_2 A_1}\right)\left(\frac{\partial_r A_3}{A_2 A_3}\right)
    \nonumber\\
  &
  +4\left(\frac{\partial_t A_2}{A_1 A_2}\right)\left(\frac{\partial_t A_3}{A_1 A_3}\right)
  -\frac{2}{A_3^2}.
\end{align}
The symmetric part of the field equations are
\begin{subequations}\allowdisplaybreaks
\begin{align}\label{Symmetric_FE}
\kappa^2 \rho&=-\frac{1}{2}\left(f(T)-Tf'(T)\vphantom{\frac{1}{A_3^2}}\right)
-2f''(T)\left(\frac{\partial_r T}{A_2}\right)\left(\frac{\cos(\chi)\cosh(\psi)}{A_3}+\frac{\partial_r A_3}{A_2A_3}\right) \nonumber\\
&\qquad +f'(T)\biggl[-2\frac{\partial_r^2 A_3}{A_2^2A_3}+2\left(\frac{\partial_r A_2}{A_2^2}\right)\left(\frac{\partial_r A_3}{A_2 A_3}\right)
-\left(\frac{\partial_r A_3}{A_2A_3}\right)^2+\frac{1}{A_3^2}\nonumber \\
&\qquad\qquad\qquad+2\left(\frac{\partial_t A_2}{A_1 A_2}\right)\left(\frac{\partial_t A_3}{A_1 A_3}\right)+\left(\frac{\partial_t A_3}{A_1 A_3}\right)^2\biggr] , \\
 0&=-f''(T)\left[\left(\frac{\partial_r T}{A_2}\right)\left(\frac{\cos(\chi)\sinh(\psi)}{A_3}+\frac{\partial_t A_3}{A_1A_3}\right)+\left(\frac{\partial_t T}{A_2}\right)\left(\frac{\cos(\chi)\cosh(\psi)}{A_3}+\frac{\partial_r A_3}{A_2A_3}\right)\right]
\nonumber \\
&\quad\ +f'(T)\biggl[-2\frac{\partial_t\partial_r A_3}{A_1A_2A_3}
+2\left(\frac{\partial_r A_1}{A_1A_2}\right)\left(\frac{\partial_t A_3}{A_1 A_3}\right)
+2\left(\frac{\partial_r A_3}{A_2A_3}\right)\left(\frac{\partial_t A_2}{A_1A_2}\right)\biggr],
\\
\kappa^2 P_r&=\frac{1}{2}\left(f(T)-Tf'(T)\vphantom{\frac{1}{A_3^2}}\right)
-2f''(T)\left(\frac{\partial_t T}{A_1}\right)\left(\frac{\cos(\chi)\sinh(\psi)}{A_3}+\frac{\partial_t A_3}{A_1A_3}\right)&\nonumber \\
&\qquad +f'(T)\biggl[
2\left(\frac{\partial_r A_1}{A_2A_1}\right)\left(\frac{\partial_r A_3}{A_2 A_3}\right)
+\left(\frac{\partial_r A_3}{A_2A_3}\right)^2
-\frac{1}{A_3^2}\nonumber\\
&\qquad\qquad\qquad  +2\left(\frac{\partial_t A_1}{A_2A_1}\right)\left(\frac{\partial_t A_3}{A_1 A_3}\right)
-\left(\frac{\partial_t A_3}{A_1A_3}\right)^2
-2\left(\frac{\partial_t^2 A_3}{A_1^2A_3}\right)
\biggr],\\
\nonumber \\ 
\kappa^2 P_t&=\frac{1}{2}\left(f(T)-Tf'(T)\vphantom{\frac{1}{A_3^2}}\right)\nonumber \\
&\qquad +f''(T)\biggl[
\left(\frac{\partial_r T}{A_2}\right)\left(\frac{\cos(\chi)\cosh(\psi)}{A_3} -\frac{\partial_t \psi}{A_1}+\frac{\partial_r A_1}{A_1A_2}+\frac{\partial_r A_3}{A_2A_3}\right)&\nonumber \\
&\qquad\qquad\qquad+ \left(\frac{\partial_t T}{A_1}\right)
\bigg(-\frac{\cos(\chi)\sinh(\psi)}{A_3}+\frac{\partial_r \psi}{A_2}-\frac{\partial_t A_3}{A_1A_3}
 -\frac{\partial_t A_2}{A_1A_2}\bigg)\biggr]&\nonumber\\
&\quad+f'(T)\biggl[\frac{\partial_r^2 A_1}{A_2^2A_1} + \frac{\partial_r^2 A_3}{A_2^2A_3}
-\left(\frac{\partial_r A_1}{A_2A_1}\right)\left(\frac{\partial_r A_2}{A_2^2}\right)
+\left(\frac{\partial_r A_1}{A_2A_1}\right)\left(\frac{\partial_r A_3}{A_2A_3}\right)
-\left(\frac{\partial_r A_2}{A_2^2}\right)\left(\frac{\partial_r A_3}{A_2 A_3}\right)
&\nonumber \\
&\quad\qquad
-\frac{\partial_t^2 A_2}{A_1^2A_2}
-\frac{\partial_t^2 A_3}{A_1^2A_3}
+\left(\frac{\partial_t A_1}{A_1^2}\right)\left(\frac{\partial_t A_2}{A_1A_2}\right)
+\left(\frac{\partial_t A_1}{A_1^2}\right)\left(\frac{\partial_t A_3}{A_1A_3}\right)
-\left(\frac{\partial_t A_2}{A_1A_2}\right)\left(\frac{\partial_t A_3}{A_1A_3}\right)
\biggr].
\end{align}
\end{subequations}
The antisymmetric part of the field equations yield
\begin{subequations}
\begin{eqnarray}
0 &=& f''(T)\left[\left(\frac{\partial_r T}{A_2}\right) \left(\frac{\cos(\chi)\,\sinh(\psi)}{A_3} +\frac{\partial_t A_3}{A_1 A_3}\right)-\left(\frac{\partial_t T}{A_1}\right) \left(\frac{\cos(\chi)\,\cosh(\psi)}{A_3}+\frac{\partial_r A_3}{A_2A_3}\right)\right],\quad
\\
0 &=&f''(T)\,\frac{\sin(\chi)}{A_3}\,\left[\left(\frac{\partial_r\,T}{A_2}\right) \cosh(\psi)-\left(\frac{\partial_t\,T}{A_1}\right)\sinh(\psi)\right].
\end{eqnarray}
\end{subequations}
The Noether identity associated with the principle of equivalence, yields the energy-momentum conservation equations
\begin{subequations}
\begin{eqnarray}
0 &=& \frac{\partial_t{A_2}}{A_2}(\rho+P_r) +2\frac{\partial_t A_3}{A_3} (\rho+P_t) +\partial_t \rho ,
\\
0 &=& \frac{\partial_rA_1}{A_1}(\rho+P_r) +2\frac{\partial_rA_3}{A_3}(P_r-P_t) +\partial_r P_r.
\end{eqnarray}
\end{subequations}

\subsection{Proper Orthonormal Gauge or the Diagonal Orthonormal Gauge?}

Above we expressed the field equations for $f(T)$ teleparallel gravity in the Diagonal Orthonormal gauge.  Which co-frame/spin connection pair should one use in computations in spherically symmetric teleparallel theories of gravity?  To illustrate the effectiveness of using the diagonal co-frame \eqref{diagonal_frame} and corresponding spin connection \eqref{orthonormal_connection} with the proper co-frame \eqref{SS_proper_frame} we compute the first component of the torsion for each. In the proper co-frame \eqref{SS_proper_frame} we have
\begin{eqnarray}
\pT^1&=&\left[
  \left(-\frac{\partial_r \psi}{A_2} + \frac{\partial_t A_2}{A_1A_2}\right)\sinh(\psi)
+ \left( \frac{\partial_t \psi}{A_1} - \frac{\partial_r A_2}{A_1A_2}\right)\cosh(\psi)\right] \nonumber\\
&&\qquad \times \biggl(\cos(\theta)\,\ph^1 \wedge \ph^2 - \cos(\phi)\sin(\theta)\,\ph^1 \wedge \ph^3 -\sin(\phi)\sin(\theta)\,\ph^1\wedge \ph^4\biggr) .\nonumber
\end{eqnarray}
Alternatively, employing the diagonal co-frame \eqref{diagonal_frame} and spin connection \eqref{orthonormal_connection} the first component of the torsion is
\begin{equation}
T^1 = \left(\frac{\partial_t \psi}{A_1}-\frac{\partial_r A_1}{A_1A_2}\right)\,h^1 \wedge h^2 +\frac{2}{A_3}\sinh(\psi)\sin(\chi)\,h^3 \wedge h^4.
\end{equation}
Of course the expressions for $\pT^a$ and $T^a$ are related through the Lorentz transformation \eqref{Lorentz-Matrix} and therefore represent equivalent geometrical objects.  The expressions for the other torsion terms computed in the proper frame are even lengthier.

We observe, that the choice of gauge, Diagonal Orthonormal versus Proper Orthonormal is of the utmost importance when investigating teleparallel theories of gravity, including but not limited to $f(T)$ teleparallel gravity.  While we have complete freedom to choose any co-frame and its corresponding spin connection, one choice may be easier to use than the other.  While the spin connection is zero in the Proper Orthonormal gauge, the equations for the torsion and field equations are significantly more complicated.  Whereas, in the Diagonal Orthonormal gauge, the torsion and the field equations are more tractable, and may yield  insights into dynamics that may be obfuscated in the Proper Orthonormal gauge.


\section{Special Subcases in \texorpdfstring{$f(T)$}{f(T)} Teleparallel gravity}

\subsection{Static Spherically Symmetric Geometries in \texorpdfstring{$f(T)$}{f(T)} gravity}

From a general spherically symmetric geometry there are various important special subcases that arise naturally from assuming an additional affine frame symmetry. For example, static spherically symmetric geometries can be generated by assuming an additional affine frame symmetry generator of the form
\begin{equation}
{\bf X}_{(4)}=\partial_t.
\end{equation}
These particular gravitational models in $f(T)$ teleparallel gravity have been widely studied in the literature \cite{Ferraro:2011ks,HamaniDaouda:2011iy,Wang:2011xf,Krssak_Pereira2015,Landry:2024dzq,Ruggiero:2015oka,Krssak_Saridakis2015,
DeBenedictis:2016aze,Bahamonde:2021gfp,Coley:2024tqe} typically in vacuum but not always in the covariant version of the theory.

\subsection{Kantowski-Sachs Spatially Homogeneous Cosmological Models in \texorpdfstring{$f(T)$}{f(T)} gravity} \label{Kantowski-Sachs}

An alternative to the static assumption above is an assumption where the additional affine frame symmetry generator is of the form
\begin{equation}
{\bf X}_{(4)}=\partial_r.
\end{equation}
This symmetry assumption yields the important spatially homogenous and anisotropic Kantowski-Sachs geometry that is used as an anisotropic generalization of the positive cuvature Robertson-Walker cosmological models in GR. In this case, the five arbitrary functions all become functions strictly of time.    Further, as explained earlier, we can easily define a new time coordinate $t'$ so that $dt'=A_1(t)dt$, thereby reducing the number of arbitrary functions from five to four without loss of generality.  Most easily done by setting $A_1(t)=1$ in equation \eqref{diagonal_frame}.

In \cite{vandenHoogen:2023pjs}, it was shown that not all spatially homogeneous and anisotropic geometries are consistent with the $f(T)$ teleparallel gravity field equations unless $f(T)=T$ or $T$ is a constant.  For Bianchi type B geometries which include Bianchi types $IV$, $V$, $VI_h$ and $VII_h$ there is no solution to the antisymmetric part of the field equations that respects the symmetry requirements unless $f(T)$ is linear in $T$ or $T$ is a constant.  Alternatively, for Bianchi type A geometries, there always exists a solution to the antisymmetric part of the $f(T)$ teleparallel gravity field equations that does not require $f(T)$ be linear in $T$ or $T$ be a constant.  To complete the consistency check of all the spatially homogenous and anisotropic models in $f(T)$ teleparallel gravity, requires that we analyze the antisymmetric part of the field equations for the Kantowski-Sachs geometries.  If we can choose a $\psi(t)$ and $\chi(t)$ that satisfies that antisymmetric part of the $f(T)$ teleparallel gravity field equations, then we can add the Kantowski-Sachs geometries to the Bianchi type A geometries which all can be used to find non-trivial, anisotropic $f(T)$ teleparallel cosmological solutions.

For the Kantoswki-Sachs geometry assuming a co-moving perfect fluid with energy density $\rho$, and isotropic pressure $P=P_t=P_r$, the antisymmetric part of the field equations are
\begin{subequations}\label{KS Asym FEs}
\begin{align}
    \frac{1}{ A_{3}(t)}f''(T) \partial_t{T} \cos(\chi) \cosh(\psi) &=0, \\
    \frac{1}{ A_{3}(t)}f''(T) \partial_t{T} \sin(\chi) \sinh(\psi) &=0,
\end{align}
\end{subequations}
where we observe that a solution to the antisymmetric part of the field equations is $\psi(t) = 0 $ and $\chi(t)=(n+\frac{1}{2})\pi$ for $ n \in \mathbb{N}$. To include solutions for all values of $\chi$ we introduce the discrete parameter $\delta$, where $\sin(\chi) = \delta = \pm 1$.  One easily concludes that it is possible to find a  solution to the antisymmetric part of the $f(T)$ teleparallel gravity field equations if the geometry is Kantowski-Sachs.  That is the assumption of a Kantowski-Sachs geometry is indeed consistent with the $f(T)$ teleparallel gravity field equations.

Assuming that $\psi = 0 $, $\cos(\chi)=0$ and $\sin(\chi)=\delta$, the torsion scalar is
\begin{equation}\label{KSTscal}
T = 2\left(\frac{\partial_t A_3}{A_3}\right)^2
  +4\left(\frac{\partial_t A_2}{A_2}\right)\left(\frac{\partial_t A_3}{A_3}\right)
  -\frac{2}{A_3^2}.
\end{equation}
Further the symmetric part of the Kantowski-Sachs field equations are,
\begin{subequations}\label{KS sym FEs}
\begin{align}
    \kappa^2 \rho &=-\frac{1}{2}(f(T)-Tf'(T))+f'(T)\left( \frac{1}{A_{3}^2} +2\left(\frac{\partial_{t}A_{2}}{A_{2}}\right)\left(\frac{\partial_{t}A_{3}}{ A_{3}}\right) +\left(\frac{\partial_{t}A_{3}}{ A_{3}}\right)^2\right), \\
    \kappa^2 P &= \frac{1}{2}(f(T)-Tf'(T))-2f''(T)\partial_t T \left(\frac{\partial_t A_3}{A_3}\right) \nonumber \\  & \qquad  \qquad \qquad \qquad \qquad \qquad  -f'(T) \left(2\left(\frac{\partial^{2}_t A_3}{A_3}\right)+\left(\frac{\partial_t A_3}{A_3}\right)^2+ \frac{1}{A^{2}_{3}}\right), \\
    \kappa^2 P &= \frac{1}{2}(f(T)-Tf'(T))-f''(T)\partial_t T \left(\frac{\partial_t A_3}{A_3} +\frac{\partial_t A_2}{A_2} \right) \nonumber \\  & \qquad \qquad \qquad \qquad \qquad  -f'(T) \left(\left(\frac{\partial^{2}_t A_2}{A_2}\right)+\left(\frac{\partial^{2}_{t} A_3}{A_3}\right) +\left(\frac{\partial_t A_2}{A_2}\right)\left( \frac{\partial_t A_3}{A_3}\right)\right).
\end{align}
\end{subequations}
Along with the associated energy momentum conservation equation
\begin{equation}
    0=\left(\rho +P\right)\left(2\frac{\partial_t A_3}{A_3} +\frac{\partial_t A_2}{A_2}\right) +\partial_t \rho
\end{equation}
We observe that the torsion scalar and the symmetric part of the field equations are independent of the parameter $\delta$.

\subsection{Power-law Solutions to the Vacuum Field Equations}

In the vacuum case we assume $\rho(t)=P(t)=0$, we will also choose $\psi(t) = 0 $ and $\chi(t)=(n+\frac{1}{2})\pi$ for $ n \in \mathbb{N}$ to solve the antisymmetric field equations identically. Left with only the symmetric equations to solve, various choices of ansatzes can be made, either to solve for an expression for $f(T)$ or find expressions for $A_{2}(t) $ and $A_{3}(t)$.

\subsubsection{Power-Law Ansatz}
Assuming $A_2(t)=t^{\alpha_{2}}$ and $A_{3}(t)=c_{3} t^{\alpha_{3}}$ the torsion scalar, \eqref{KSTscal} becomes,
\begin{equation}\label{KSpowerlawTscal}
    T = \frac{2\alpha_{3}(\alpha_{3}+2\alpha_{2})}{t^{2}}-\frac{2}{c^{2}_{3}t^{2\alpha_{3}}}.
\end{equation}
The Kantowski-Sachs symmetric field equations, \eqref{KS sym FEs} in vacuum can be expressed as,
\begin{subequations}
\begin{align}
    f(T) &=2\left(T+\frac{2}{c^{2}_{3} t^{2 \alpha_{3}}}\right)f'(T), \label{KSpowerlawFEs:1} \\
    tf''(T) &= \left(1-\alpha_{2}-2\alpha_{3}\right) + \frac{t^{2-2\alpha_{3}}}{c^{2}_{3}\left(\alpha_{2}-\alpha_{3}\right)}, \label{KSpowerlawFEs:2} \\
    0 &= \alpha_{3} \left(tf''(T) +\left(\alpha_{3}-\alpha_{2}-1\right)\right).\label{KSpowerlawFEs:3}
\end{align}
\end{subequations}
Substituting \eqref{KSpowerlawFEs:2} into \eqref{KSpowerlawFEs:3} we find the following consistency equation,
\begin{equation}\label{KSpowerlawconsistencyeq}
    0=-c^{2}_{3}\left(\alpha_{2}-\alpha_{3}\right)\left(2\alpha_{2}+\alpha_{3}\right) +t^{2-2\alpha_{3}}.
\end{equation}
We need this equation to be satisfied for all values of $t$ so $\alpha_{3}= 1$ and
\begin{equation}
c_3^2= \frac{1}{(2\alpha_2+1)(\alpha_2-1)},
\end{equation}
where $\alpha_2< -\nicefrac{1}{2}$  or $\alpha_2>1$ yields real valued solutions.
An explicit expression for $f(T)$ can be found by solving \eqref{KSpowerlawTscal} and \eqref{KSpowerlawFEs:1} yielding
\begin{equation}
f(T)=f_{0} T^{1-\frac{\alpha_2}{2}}, \mbox{\ where\ } T= \frac{2(1+2\alpha_2)(2-\alpha_2)}{t^2},
\end{equation}
where $f_0$ is an arbitrary constant of integration.  If we require $T(t)>0$ to ensure a non-trivial real valued function $f$ then the value of the parameter $\alpha_2$ is restricted to the interval $1<\alpha_2<2$. If we define quantities which resemble the usual expansion $\theta(t)$ and anisotropy $\sigma(t)$ to be
\begin{subequations} \label{def_H_Sigma}
\begin{align}
\theta(t)&:=\frac{\partial_t{A_2}}{A_2}+2\frac{\partial_t{A_3}}{A_3},\\
\sigma(t)&:= \frac{\partial_t{A_2}}{A_2}-\frac{\partial_t{A_3}}{A_3},
\end{align}
\end{subequations}
then we observe for this solution that $\theta(t)=(\alpha_2+2)t^{-1}$ and $\sigma(t)=(\alpha_2-1)t^{-1}$.  This represents an expanding anisotropic vacuum solution which isotropizes as $t\to\infty$. This solution without the restrictions on the parameters and additional solutions for Kantowski-Sachs geometries in $f(T)$ teleparallel gravity are found in \cite{Landry:2024pxm}.

\subsubsection{\texorpdfstring{$f(T)=T+\varepsilon T^2$}{f(T)=T+epsilon T2}  Ansatz}

Another approach involves using an ansatz for $f(T)$ and working backwards to find $A_{2}(t)$ and $A_{3}(t)$. We want to find solutions that are approximately TEGR ($f(T)=T$) with a small quadratic addition, resulting in the ansatz $f(T)=T+\varepsilon T^2$. Subbing the ansatz for $f(T)$ into \eqref{KS sym FEs} gives a set of differential equations which can be solved yielding the following solutions,
\begin{equation}
 A_{2}(t) = \cosh{\left(\frac{t}{\sqrt{2 \varepsilon}}\right)} \ \ \ \ \ \ \ \ \ \ \     A_{3}(t)=\sqrt{2 \varepsilon},
\end{equation}
where we observe that $\epsilon>0$.  The result of $A_{3}(t)$ being constant gives a constant torsion scalar,
\[T= -\frac{1}{\varepsilon}.\]
thus, this solution is not a $f(T)$ teleparallel gravity solution since $f(T)$ is identically zero indicating that the ansatz $f(T)=T+\varepsilon T^2$ is not compatible with the geometry.

\subsection{Power-law Solutions to the Perfect Isotropic Fluid Field Equations}

As before, we will choose $\psi(t) = 0 $ and $\chi(t)=(n+\frac{1}{2})\pi$ for $ n \in \mathbb{N}$ to solve the antisymmetric field equations identically.  We assume a perfect isotropic fluid where $P(t)=(\gamma-1)\rho(t)$ with equation of state parameter $0<\gamma\leq 2$.  In contrast to what was done earlier, to eliminate one of the frame functions, we will choose coordinates so that $A_3(t)=t$ leaving only $A_1(t)$ and $A_2(t)$. As before we seek simple power-law solutions of the form $A_1(t)=c_1t^{\alpha_1}$, $A_2=c_2t^{\alpha_2}$.  The value of $c_2$ can be set to $1$ via a coordinate change without loss of generality. Assuming that $T\propto t^{-2}$ implies that $\alpha_1=0$.  The conservation and field equations can easily be solved to yield
\begin{equation}
\rho(t)=\rho_0 t^{-2\tau}, \qquad  f(T)=f_0 \left(\frac{T}{T_0} \right)^\tau , \mbox{\ \  and\ \ }  T=\frac{T_0}{t^2},
\end{equation}
where
\begin{subequations}
\begin{eqnarray}
\tau &=& \frac{\gamma}{2}(\alpha_2+2),\\
T_0 &=& \left(\frac{2}{c_1^2}(2\alpha_2+1)-2\right),\\
f_0&=&2\kappa\rho_0 \left(\frac{\gamma(\alpha_2-1)-(\alpha_2-2)}{\gamma(\alpha_2+2)+(\alpha_2-2)}\right),\\
c_1^2&=&(\alpha_2-1)\bigl( (\alpha_2+3) - \gamma(\alpha_2+2)\bigr),
\end{eqnarray}
\end{subequations}
and $\rho_0$   is an  arbitrary constant of integration.  In order to have real nontrivial solutions for $c_1$ and requiring $T(t)>0$ to have a real valued function $f$, the following inequality
\begin{equation} \label{ineq}
0<c_1^2=(\alpha_2-1)\bigl( (\alpha_2+3) - \gamma(\alpha_2+2)\bigr) < 1+2\alpha_2,
\end{equation}
restricts the possible values for $\gamma$ and $\alpha_2$. For example in the dust case when $\gamma=1$ inequality \eqref{ineq} implies that $\alpha_2>1$ provides a valid power-law dust solution.
The expansion $\theta(t)$ and anisotropy $\sigma(t)$ due to the change of coordinates are given by
\begin{subequations} \label{def_H_Sigma2}
\begin{align}
\theta(t)&:=\frac{1}{A_1}\left(\frac{\partial_t{A_2}}{A_2}+2\frac{\partial_t{A_3}}{A_3}\right),\\
\sigma(t)&:=\frac{1}{A_1}\left(\frac{\partial_t{A_2}}{A_2}-\frac{\partial_t{A_3}}{A_3}\right),
\end{align}
\end{subequations}
where it is understood that $A_3=t$ in the chosen coordinates. We find again that $\theta(t)=(\alpha_2+2)t^{-1}$ and $\sigma(t)=(\alpha_2-1)t^{-1}$.
This two-parameter $\{\gamma,\alpha_2\}$ solution satisfying the inequality \eqref{ineq} with $0<\gamma\leq 2$ gives a family  of special expanding anisotropic power-law solutions to the Kantowski-Sachs perfect fluid field equations in $f(T)$ teleparallel gravity.


\section{Concluding remarks}

The diagonal orthonormal co-frame \eqref{diagonal_frame} and corresponding spin connection \eqref{orthonormal_connection} determined in \cite{McNutt:2023nxm} is one of the simplest ansatzes to employ in the study of spherically symmetric teleparallel geometries. This diagonal co-frame/spin connection pair has five arbitrary functions of the time and radial coordinates, two of which are found in the spin connection. It is always possible to reduce the number of arbitrary functions to four, although one necessarily constructs a non-diagonal co-frame in the process.  It is of some interest to transform this diagonal orthonormal frame to a proper (non-diagonal) orthonormal frame in which the spin connection is trivial.  With this co-frame choice, there are no inertial effects in the frame.  We have shown how the resulting Lorentz transformation relating the diagonal frame and the proper frame is a composition of a number of Rotations and Boosts dependent on two functions of the time and radial coordinates found in the spin connection.  We also explicitly computed the proper orthonormal co-frame and compared it to other teleparallel spherically symmetric proper co-frames in the literature.

We compared the expressions for the torsion in both the diagonal orthonormal gauge and in the proper orthonormal gauge.  We easily see that it may be much easier to obtain results using the diagonal orthonormal gauge with the non-trivial spin connection than it is in the proper orthonormal gauge with a trivial spin connection.  Of course the results are identical in the sense that the torsion and field equations in one frame can be transformed into the torsion and field equations of the other frame simply by applying the Lorentz transformation. But this only works if you actually know the Lorentz transformation that relates the diagonal frame to the proper frame.

Further, in the spatially homogenous but anisotropic Kantowski-Sachs geometry, it was shown there exists a spin connection that satisfies the antisymmetric part of the $f(T)$ teleparallel gravity field equations.   This means that the Kantowski-Sachs teleparallel geometries are consistent with $f(T)$ teleparallel gravity and belong to the same class of consistent spatially homogeneous and anisotropic geometries as the Bianchi type A geometries.  The Kantowski-Sachs geometry was the last spatially homogeneous and anisotropic cosmological model to be analyzed in this way, thereby completing the analysis in \cite{vandenHoogen:2023pjs}. In the parlance of the research community, there exists a ``good tetrad'' for spatially homogeneous but anisotropic geometries of Bianchi type A and Kantowski-Sachs geometries.

The exposition presented here clarifies the nature of the non-trivial spin connection used in spherically symmetric teleparallel geometries.  Indeed, while the field equations and analysis of the Kantowski-Sachs geometries was completed in the $f(T)$ class of teleparallel gravity theories, the next step is to naturally apply these same techniques to other classes of teleparallel gravity theories.


\begin{acknowledgments}
RvdH was supported by the Natural Sciences and Engineering Research Council of Canada and by the W.F. James Chair of Studies in the Pure and Applied Sciences at St.F.X. HF was supported by the Natural Sciences and Engineering Research Council of Canada. The authors would like to thank Alan Coley, Alexandre Landry and Diego L\'{o}pez for their comments on earlier versions of this paper.
\end{acknowledgments}



\bibliographystyle{JHEP}

\providecommand{\href}[2]{#2}\begingroup\raggedright\endgroup

\end{document}